\documentclass[dvipsnames]{article}

\usepackage[final]{openbmb_2025}

\usepackage{amsmath}
\usepackage{amssymb}
\usepackage{mathtools}
\usepackage{amsthm}
\usepackage{multirow}
\usepackage{makecell}
\usepackage{subcaption}
\usepackage{wrapfig}
\usepackage{graphicx}

\usepackage{pifont}
\usepackage{fvextra}
\usepackage{xspace}

\usepackage[utf8]{inputenc} 
\usepackage[T1]{fontenc}    
\usepackage{url}            
\usepackage{booktabs}       
\usepackage{amsfonts}       
\usepackage{nicefrac}       
\usepackage{microtype}      

\usepackage{color, soul}
\usepackage[most,breakable]{tcolorbox}
\usepackage{textgreek}

\usepackage[labelfont=bf]{caption}
\usepackage{enumitem}
\usepackage{sectsty}
\usepackage{pgfplots}
\usepackage{fancyhdr}
\usepackage{xcolor} 
\usepackage{bbding}
\renewcommand{\headrulewidth}{1pt}
\makeatletter
\def\headrule{{\if@fancyplain\let\headrulewidth\plainheadrulewidth\fi
\hrule\@height\headrulewidth\@width\textwidth \vskip-\headrulewidth}}
\makeatother

\definecolor{BMBDarkBlue}{HTML}{315EFE}
\definecolor{BMBLightBlue}{HTML}{00D3ED}
\usepackage[colorlinks, linkcolor=BMBDarkBlue, anchorcolor=BMBDarkBlue, citecolor=BMBDarkBlue, urlcolor=BMBDarkBlue]{hyperref}
\usepackage{cleveref}
\usepackage{threeparttable}

\sectionfont{\color{BMBDarkBlue}\fontfamily{zi4}\selectfont}
\subsectionfont{\color{BMBDarkBlue}\fontfamily{zi4}\selectfont}
\subsubsectionfont{\color{BMBDarkBlue}\fontfamily{zi4}\selectfont}

\newtcolorbox{mytheorem}{
  colback=gray!5,       
  colframe=gray!80,     
  boxrule=0.5pt,        
  arc=4pt,              
  left=4pt,             
  right=4pt,            
  top=4pt,              
  bottom=4pt,           
}

\newcommand{\fancyheadname}{\textit{\textbf{\modelname{}}}}

\title{VoxCPM2 Technical Report}

\author{%
\\
\textbf{\large{VoxCPM Team}}
\vspace{0em}
}

\def\huggingface{\raisebox{-1.5pt}{\includegraphics[height=1.05em]{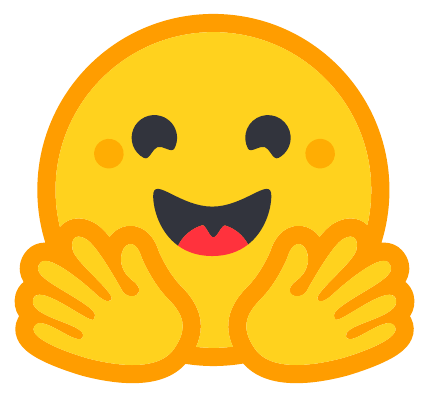}}}
\def\github{\raisebox{-1.5pt}{\includegraphics[height=1.05em]{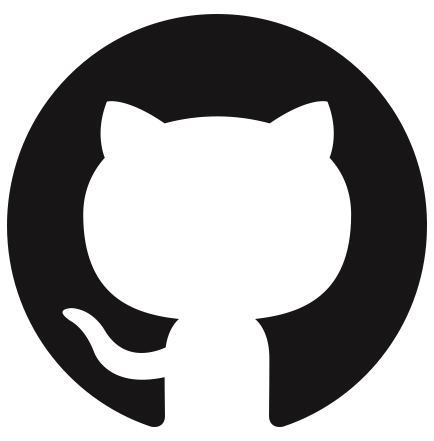}}}
\newcommand{\modelname}[0]{VoxCPM2}

\usepackage{geometry}
\usepackage{algorithm}
\usepackage{algpseudocode}
\usepackage{setspace}

\makeatletter
\renewcommand{\ALG@beginalgorithmic}{\small}
\makeatother



\begin{document}
\maketitle
\thispagestyle{fancy}
\begin{abstract}
We present \modelname{}, a fully open-source multilingual and controllable speech generation foundation model that extends the hierarchical diffusion-autoregressive modeling paradigm of VoxCPM.
\modelname{} advances the framework in three key dimensions:
(i) \textbf{capability}, by unifying 30 languages, 9 Chinese dialects, natural-language voice design, style-controllable voice cloning, and high-fidelity continuation cloning within a single backbone;
(ii) \textbf{quality}, through an asymmetric AudioVAE that encodes at 16 kHz and reconstructs at 48 kHz, enabling implicit super-resolution with high encoding efficiency;
and (iii) \textbf{scale}, by jointly scaling the model to 2B parameters and the training data to over 2 million hours of multilingual speech.
To support these diverse capabilities within one model, we introduce a unified sequence organization that expresses all generation modes through different arrangements of the same input building blocks, allowing joint training under a single set of parameters and objective.
\modelname{} achieves state-of-the-art or competitive performance on public zero-shot and instruction-following TTS benchmarks. 
On our internal 30-language evaluation set, it attains an average WER of 1.68\%.
These results demonstrate that hierarchical continuous-latent modeling, without relying on any external discrete speech tokenizer, offers a viable and powerful foundation for large-scale multilingual and controllable speech generation.
The model weights, fine-tuning code, and inference tools are publicly released under the Apache 2.0 license to foster community research and development.

\end{abstract}

\newpage
{
  \hypersetup{linkcolor=RoyalBlue, linktoc=page}
  \tableofcontents
}

\newpage

\section{Introduction}
\label{sec:intro}
\subsection{Background and VoxCPM Foundation}

Text-to-speech (TTS) has evolved from producing intelligible speech toward generating natural, expressive, and controllable audio~\citep{ping2017deep, shen2018natural, renfastspeech, li2019neural}. 
Modern applications—such as conversational agents, dubbing, accessibility tools, and interactive digital characters—require not only accurate pronunciation but also faithful reproduction of speaker identity, speaking style, and communicative intent. This raises the bar for acoustic fidelity, controllability, and multilingual coverage.

Driven by the success of large language models (LLMs), the dominant paradigm in contemporary TTS frames speech synthesis as sequence modeling over discrete tokens produced by neural audio codecs or tokenizers~\citep{defossez2022high, kumar2023high, zhang2024speechtokenizer}. 
By discretizing speech, these systems inherit the scaling laws and in-context learning capabilities of LLMs~\citep{borsos2023audiolm, kharitonov2023speak, chen2025neural}.
Recent advances have further extended this paradigm through improved tokenizer designs, fine-grained prosody and emotion control, and multilingual long-form generation~\citep{peng2024voicecraft, wang2025spark, wangmaskgct, hu2026qwen3, gong2026moss, liao2026fish}. 

However, quantization inevitably discards fine-grained acoustic details. 
To mitigate this loss, most high-quality token-based systems adopt multi-stage pipelines, in which an autoregressive LLM predicts coarse or semantic tokens while a separate diffusion or flow-matching model restores local acoustic fidelity~\citep{du2024cosyvoice1, du2024cosyvoice2, du2025cosyvoice3, zhou2025indextts2, casanova2024xtts, guo2024fireredtts, xie2025fireredtts}.
Although effective in achieving strong perceptual quality, this decoupled design fragments high-level semantic planning from low-level acoustic rendering, preventing end-to-end joint optimization.
Moreover, overall performance remains heavily dependent on the modeling capacity of the intermediate discrete speech tokenizer.

An alternative research direction models continuous speech representations directly to preserve richer acoustic information.
Building upon early autoregressive mel-spectrogram systems~\citep{shen2018natural, meng2024autoregressive}, recent methods employ denoising or flow-matching objectives over continuous acoustic latents.
These approaches include both non-autoregressive diffusion models~\citep{shen2023naturalspeech, le2023voicebox, chen2024f5} and diffusion-autoregressive hybrids~\citep{li2024autoregressive, jia2025ditar, peng2025vibevoice, wu2025clear, turetzky2025speech}. 
While effective at capturing fine acoustic details, they must jointly optimize semantic-prosodic structure and local acoustic texture within the same representation space and training objective, often leading to optimization challenges and error accumulation in long-form or highly expressive generation.

VoxCPM~\citep{zhou2025voxcpm} was proposed to address this fundamental trade-off. 
It introduces a hierarchical backbone consisting of a Text-Semantic Language Model (TSLM), a differentiable semi-discrete bottleneck based on Finite Scalar Quantization (FSQ)~\citep{mentzerfinite}, and a Residual Acoustic Language Model (RALM).
The TSLM primarily captures high-level semantic and prosodic structures, the FSQ bottleneck compresses them into a stable skeleton, and the RALM recovers fine-grained acoustic details. These components jointly condition a Local Diffusion Transformer (LocDiT) to generate continuous latent patches.
By leveraging this internal hierarchical design with a differentiable semi-discrete bottleneck, VoxCPM enables end-to-end training on continuous latents without any external discrete speech tokenizer. 
This structure facilitates joint optimization of semantic planning and acoustic rendering while mitigating the fragmentation typical of multi-stage pipelines. 

Overall, VoxCPM demonstrates that hierarchical continuous-latent modeling can achieve competitive performance without sacrificing acoustic richness or semantic intelligibility.
Building on this foundation, \modelname{} evolves VoxCPM into a strong and practical TTS foundation model by significantly advancing capability, quality, and scale, while strictly preserving the hierarchical end-to-end continuous-latent design.

\subsection{VoxCPM2: From Foundation to Full-Featured System}

\modelname{} is the latest major release in the VoxCPM family—a 2B-parameter hierarchical diffusion-autoregressive speech generation model built upon the MiniCPM-4 backbone~\citep{team2025minicpm4}. 
It advances the original framework along three core dimensions: capability, quality, and scale, while preserving the hierarchical continuous-latent design.

\textbf{Capability.} \modelname{} unifies \textbf{four user-facing capabilities} within a single backbone: 
(i) basic TTS supporting multilingual and cross-lingual synthesis, 
(ii) natural-language voice design that generates entirely new voices from free-form text descriptions without any reference audio,
(iii) controllable cloning that clones a speaker from a short reference audio while following style instructions, 
and (iv) continuation-based cloning for high-fidelity audio continuation from a paired reference audio and its transcript.
All modes share the same parameters, training objective, and inference pipeline, differing only in input sequence organization.

\textbf{Quality.} We introduce AudioVAE V2, an asymmetric latent codec that encodes at the sampling rate of 16~kHz and reconstructs at \textbf{48~kHz}. This design maintains compact latent sequences for the \modelname{} backbone while enabling implicit super-resolution and high-quality output.

\textbf{Scale.} Recent work has formalized the Densing Law of LLMs~\citep{xiao2025densing}, which shows that the capacity density of LLMs (effective performance per parameter) grows exponentially, roughly doubling every three months. Guided by this principle, we jointly scale \modelname{} to \textbf{2B parameters} and the training data to over \textbf{2 million hours} of multilingual speech covering 30 languages and 9 Chinese dialects, while maintaining a compact 6.25~Hz token rate.

The main contributions of \modelname{} are as follows:

\begin{enumerate}
    \item We extend the hierarchical continuous-latent framework into a unified \textbf{2B-parameter, 48~kHz, 30-language} foundation model (plus 9 Chinese dialects), while preserving end-to-end training and a compact 6.25~Hz token rate without external discrete tokenizers.
    \item We integrate \textbf{basic TTS, natural-language voice design, controllable cloning, and continuation-based cloning} into a single backbone via a unified sequence organization, replacing per-task specialized models with one set of parameters and a unified inference path.
    \item We introduce key architectural refinements—including improved semantic-acoustic fusion, multi-token LocDiT conditioning, and an isolated reference-audio pathway—to better support large-scale multilingual and controllable generation.
    \item We demonstrate strong empirical performance and practical deployability through competitive or state-of-the-art results on multiple public benchmarks, an average WER of \textbf{1.68\%} on our internal 30-language test set, and efficient streaming inference.
\end{enumerate}

\subsection{Paper Organization}

The remainder of this report is organized as follows. Section~\ref{sec:related} reviews recent progress in large-scale TTS foundation models and controllable speech generation. Section~\ref{sec:method} presents the \modelname{} system, including the overall architecture, AudioVAE V2, backbone refinements, unified sequence organization, and training strategy. 
Section~\ref{sec:exp} reports experimental results on zero-shot TTS, multilingual synthesis, controllable generation, reconstruction quality, and deployment efficiency. 
Section~\ref{sec:conclusion} discusses limitations, responsible-use considerations, and future directions.

\section{Related Work}
\label{sec:related}
We review prior work along two primary axes that frame our contributions: large-scale TTS foundation models and controllable/expressive speech generation.
\subsection{Large-Scale TTS Foundation Models}

Building on early foundation models such as AudioLM~\citep{borsos2023audiolm}, SPEAR-TTS~\citep{kharitonov2023speak}, VALL-E~\citep{chen2025neural}, and Voicebox~\citep{le2023voicebox}, recent TTS research has expanded along several complementary directions.
We organize the discussion around three paradigms most relevant to \modelname{}: discrete-token language modeling, continuous-latent generation, and hierarchical semantic-acoustic decomposition.

\paragraph{Discrete-token language modeling over neural codecs.}
The dominant paradigm represents speech as sequences of discrete tokens produced by neural audio codecs or speech tokenizers~\citep{defossez2022high, kumar2023high, xin2024bigcodec, zhang2024speechtokenizer}, inheriting LLM-style scaling and in-context learning capabilities. Three main sub-routes have emerged.

\textit{Single-backbone autoregressive systems} predict codec tokens directly with a language model. 
The dominant tokenization is residual vector quantization (RVQ), where each frame is encoded into multiple stacked codebook indices. 
While RVQ provides a richer discrete representation, it complicates joint multi-token prediction per frame.
Common strategies include coarse-to-fine prediction~\citep{borsos2023audiolm}, parallel masked prediction~\citep{borsos2023soundstorm}, and delayed or interleaved token patterns~\citep{copet2024musicgen}. 
Early systems such as VoiceCraft~\citep{peng2024voicecraft} unifies zero-shot TTS and speech editing on top of EnCodec RVQ tokens. 
Subsequent works like Llasa~\citep{ye2025llasa} explored LLM-style scaling over semantic-aware tokenizers (e.g., X-codec~\citep{ye2024xcodec}).
Complementary efforts simplify the codec interface, such as Spark-TTS~\citep{wang2025spark} with single-stream decoupled tokens and SpeechTokenizer~\citep{zhang2024speechtokenizer}, which aligns the first RVQ codebook with semantic content.
At foundation scale, models such as Qwen3-TTS~\citep{hu2026qwen3}, MOSS-TTS~\citep{gong2026moss}, Fish Audio S2~\citep{liao2026fish}, and HiggsAudio~v2~\citep{ bosonai2025higgsaudio} have demonstrated strong scaling performance.

\textit{Discrete non-autoregressive systems} replace causal autoregression with masked or parallel prediction over discrete tokens. 
SoundStorm~\citep{borsos2023soundstorm} pioneered this direction by adopting iterative masked prediction over RVQ tokens, achieving substantial speed-ups for high-fidelity audio generation while keeping a small number of refinement steps. MaskGCT~\citep{wangmaskgct} extends the idea to zero-shot TTS via a two-stage masked generative codec transformer: a semantic-stage model first predicts speech-content tokens from text, and an acoustic-stage model then predicts residual acoustic tokens conditioned on the semantic tokens, both decoded in parallel by masked generation. 
More recently, OmniVoice~\citep{zhu2026omnivoice} scales discrete masked-prediction TTS to a single multilingual model covering over 600 languages, showing that the non-autoregressive route can support large-scale multilingual coverage.

\textit{Multi-stage hybrid systems} pair an autoregressive LM for semantic or coarse-acoustic tokens with a separate diffusion or flow-matching decoder for waveform rendering.
This design has become prevalent for high perceptual quality. 
Early systems such as XTTS~\citep{casanova2024xtts} use a token-prediction LM conditioned on a speaker embedding with a HiFi-GAN-style vocoder; the CosyVoice series~\citep{du2024cosyvoice1, du2024cosyvoice2, du2025cosyvoice3} replaces the acoustic side with a flow-matching decoder conditioned on supervised semantic tokens; and the FireRedTTS series~\citep{guo2024fireredtts, xie2025fireredtts} extends this two-stage layout toward industrial-scale long-form dialogue. 
Subsequent work refines the framework for more capability or better performance: IndexTTS2~\citep{zhou2025indextts2} introduces explicit emotion and duration control, MiniMax-Speech~\citep{zhang2025minimax} learns an intrinsic speaker encoder that extracts timbre features from a reference audio without requiring its transcription, and Voxtral~TTS~\citep{liu2026voxtral} explores a hybrid VQ--FSQ codec interface that combines discrete semantic indices with continuous-valued acoustic codes. 
The same LM-plus-flow-matching pipeline has also been adopted as the speech-generation component of broader audio foundation models, including GLM-4-Voice~\citep{zeng2024glm4voice}, Step-Audio~\citep{huang2025stepaudio} and Kimi-Audio~\citep{kimiaudio2025}.

\paragraph{Continuous-latent and diffusion-autoregressive generation.}
A parallel line of research models continuous speech representations directly to preserve fine acoustic details lost during quantization. Building upon early autoregressive mel-spectrogram models~\citep{shen2018natural, meng2024autoregressive}, recent methods employ denoising or flow-matching objectives over continuous latents. 
Non-autoregressive models such as NaturalSpeech~2~\citep{shen2023naturalspeech} and Voicebox~\citep{le2023voicebox} achieve high naturalness with competitive inference speed.
MegaTTS~3~\citep{jiang2025megatts3} introduces sparse alignment to guide a latent diffusion transformer for improved handling of difficult sentences and accents.
More end-to-end approaches include E2~TTS~\citep{eskimez2024e2} and F5-TTS~\citep{chen2024f5}, which remove explicit alignment and duration modules, as well as LongCat-AudioDiT~\citep{meituan2026longcataudiodit}, which operates directly in waveform latent space using a Wav-VAE.
Diffusion-autoregressive hybrids instead couple a language model for long-range planning with a local diffusion module for fine acoustic synthesis, which offers higher expressiveness and inherits several key modeling advantages from large language models.
Notable examples include ARDiT~\citep{li2024autoregressive}, which predicts continuous mel-spectrogram frames autoregressively using a decoder-only diffusion transformer, 
and DiTAR~\citep{jia2025ditar}, which introduces a Local Diffusion Transformer (LocDiT) over patch-wise continuous latents conditioned on LM context—a foundational component reused in our work.
A series of follow-up works further extend this design~\citep{an2026mela, wang2025felle, turetzky2025speech, wu2025clear, peng2025vibevoice}.

\paragraph{Hierarchical semantic-acoustic decomposition.}
Hierarchical decomposition appears across both discrete and continuous paradigms.
The multi-stage hybrid systems implement it externally through separate semantic and acoustic stages, while other works embed hierarchy more explicitly inside the model.
HierSpeech++~\citep{lee2025hierspeechpp} bridges semantic and acoustic representations via hierarchical variational inference, HALL-E~\citep{nishimura2025halle} stacks a hierarchical neural codec with a language model to support minute-long synthesis.
and MARS6~\citep{baas2025mars6} uses a hierarchical-token encoder--decoder transformer for compact and robust generation. 
Most of these approaches rely on discrete codecs or hierarchical token vocabularies between layers.
VoxCPM~\citep{zhou2025voxcpm} takes a distinct route by realizing semantic-acoustic hierarchy inside a single continuous-latent backbone through a differentiable semi-discrete FSQ bottleneck.
This enables fully end-to-end training without external discrete tokenizers. \modelname{} scales this hierarchical continuous-latent paradigm into a large multilingual and controllable foundation model.

\subsection{Controllable and Expressive Speech Generation}  
As TTS has matured beyond intelligibility, controllability has become a central requirement—not only \textit{what} is said, but \textit{who} speaks and \textit{how}~\citep{xie2025towards}. 
Early controllable systems relied on categorical labels, global style tokens, or fixed attribute sets~\citep{wang2018style, cai2021emotion}which offered limited flexibility.
This led to the emergence of natural-language control interfaces. 
PromptTTS~\citep{guo2023prompttts} first conditions a TTS model on free-form style descriptions with BERT encoder, and PromptTTS~2~\citep{leng2024prompttts} adds variation modeling and an automatic description-generation pipeline. 
InstructTTS~\citep{yang2024instructtts} models expressive TTS in a discrete latent space conditioned on style prompts. 
For more user-friendly, PromptStyle~\citep{liu2023promptstyle} performs description-guided cross-speaker style transfer. 
Complementary latent-diffusion approaches such as VoiceLDM~\citep{lee2024voiceldm} and AudioBox~\citep{vyas2024audiobox} also investigate description-conditioned generation.

Recent advances have progressed along three main axes. 
On data construction, large-scale captioned corpora provide rich style descriptions: TextrolSpeech~\citep{ji2024textrolspeech} couples speech with style-controlling captions, SpeechCraft~\citep{jin2024speechcraft} adds fine-grained expressive annotations, and CapSpeech~\citep{wang2025capspeech} aggregates multi-source captioned speech. 
On modeling techniques, Parler-TTS~\citep{lyth2024natural} trains high-fidelity TTS conditioned on synthetic captions; VoxInstruct~\citep{zhou2024voxinstruct} unifies content and style prompts into a single instruction; and follow-up systems extend the paradigm with open-ended instructions, attribute-level editing, or reference-free voice design~\citep{yang2025emovoice, ren2026ov, hu2026voicesculptor, huang2026moss, chen2026flexivoice}.
FlexiVoice~\citep{chen2026flexivoice} and follow-ups further adopting DPO/GRPO-based post-training to better disentangle style, timbre, and content in cotrollable generation. 
Some TTS foundation systems like CosyVoice~3, Qwen3-TTS, MOSS-TTS, and Fish Audio~S2 have integrated natural-language voice generation as a native capability, with commercial platforms like Gemini TTS and ElevenLabs demonstrating production-grade performance.

Evaluation protocols have also matured in parallel. 
Benchmarks such as InstructTTSEval~\citep{huang2025instructttseval} and MINT-Bench~\citep{chen2026mint} target fine-grained adherence to natural-language instructions. 
The Audio Turing Test, used for example in the Fish Audio~S2 report~\citep{liao2026fish}, measures human-likeness via indistinguishability from real recordings.
Besides, the Timing-control Benchmark~\citep{mai2026magic} focuses on token-level duration and pause fidelity, EmergentTTS-Eval~\citep{manku2026emergenttts} probes stability under complex conditions, and TTSDS~\citep{minixhofer2024ttsds} aggregates multiple acoustic and perceptual indicators into a score, jointly marking a shift toward multidimensional and reproducible evaluation.

A common limitation in existing controllable systems is architectural fragmentation, typically involving dedicated style encoders, adapters, or per-mode routing mechanisms.
In contrast, \modelname{} treats natural-language voice and style descriptions as ordinary text prefixes to the same TSLM. 
Combined with a unified sequence organization, it supports voice design, reference-based cloning, controllable cloning, and continuation cloning within a single hierarchical continuous-latent backbone.

\section{Methodology}
\label{sec:method}

\begin{figure}[h]
\begin{center}
\includegraphics[width=1.0\textwidth]{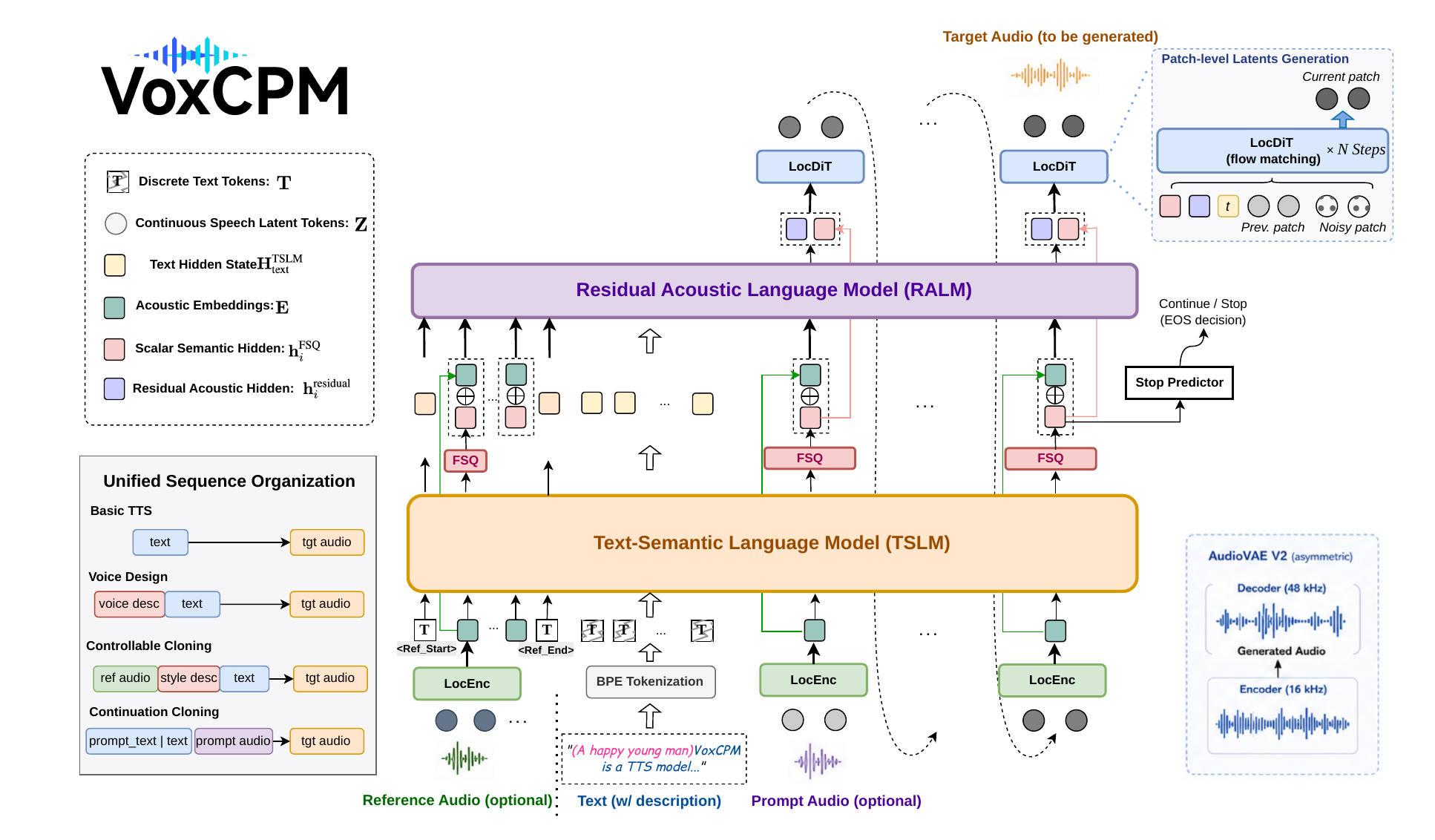}
\end{center}
\caption{Overall architecture of VoxCPM2.}
\label{fig:architecture}
\end{figure}

\subsection{Overview}
\label{sec:method-overview}
\modelname{} inherits the hierarchical diffusion-autoregressive formulation of VoxCPM~\citep{zhou2025voxcpm} and extends it into a multilingual and controllable foundation model.
Speech is modeled entirely in the continuous latent space of \textbf{AudioVAE~V2}: the encoder maps 16~kHz waveforms to 64-dimensional latent frames $z$ at 25~Hz.
The backbone then groups every $P{=}4$ frames into one \textit{patch}, resulting in a 6.25~Hz autoregressive sequence where each step corresponds to 160\,ms of audio.

The autoregressive backbone consists of a \textbf{Local Encoder (LocEnc)}, a \textbf{Text-Semantic Language Model (TSLM)}, a \textbf{Residual Acoustic Language Model (RALM)}, and a \textbf{Local Diffusion Transformer (LocDiT)},
which together predict the next continuous latent patch step by step.
Following the formulation in VoxCPM, the generation at the $i$-th patch is expressed as:
\begin{equation}
    z_i \sim \mathrm{LocDiT}\bigl(h^{\mathrm{FSQ}}_i,\; h^{\mathrm{residual}}_i,\; z_{i-1};\, t\bigr),
    \quad
    \begin{aligned}
        h^{\mathrm{FSQ}}_i &= \mathrm{FSQ}\bigl(\mathrm{TSLM}(\mathbf{T},\, \mathbf{E}_{<i})\bigr),\\
        h^{\mathrm{residual}}_i &= \mathrm{RALM}\bigl(\mathbf{H}^{\mathrm{TSLM}}_{\mathrm{text}},\; \mathbf{H}^{\mathrm{FSQ}}_{\le i} \oplus \mathbf{E}_{<i}\bigr),
    \end{aligned}
\end{equation}
where $\mathbf{T}$ denotes the input text tokens, $\mathbf{E}_{<i} = \mathrm{LocEnc}(z_{<i})$ is the patch-level acoustic history aggregated by the Local Encoder, and $t$ is the diffusion timestep. 
The FSQ layer applies per-dimension scalar quantization to the TSLM hidden states to produce the semi-discrete semantic skeleton $h^{\mathrm{FSQ}}_i$.
The RALM then recovers fine-grained acoustic details into $h^{\mathrm{residual}}_i$ by conditioning on the TSLM text-side hidden states $\mathbf{H}^{\mathrm{TSLM}}_{\mathrm{text}}$ together with a fusion ($\oplus$) of the FSQ-quantized audio-side history $\mathbf{H}^{\mathrm{FSQ}}_{\le i}$ and the Local Encoder embeddings $\mathbf{E}_{<i}$, granting causal access to the full sequence history. 
A stop predictor on top of the TSLM-FSQ hidden states determines generation termination, and the entire pipeline is trained end-to-end. 
We refer readers to the formal conference version of VoxCPM~\citep{zhou2026hierarchical} for the full derivation.

Eq.~(1) already reflects two \modelname{}-specific modifications relative to VoxCPM: 
(i) the LocDiT receives $h^{\mathrm{FSQ}}_i$ and $h^{\mathrm{residual}}_i$ as \textit{separate} conditioning tokens rather than a single summed vector $h^{\mathrm{final}}_i = h^{\mathrm{FSQ}}_i + h^{\mathrm{residual}}_i$,
and (ii) the fusion operator $\oplus$ before the RALM is replaced by a learnable concatenation-projection (detailed in Section~\ref{sec:method-backbone-architecture}). 
Three additional groups of changes transform VoxCPM into a high-fidelity, multilingual, and controllable system:
\begin{itemize}
    \item A redesigned latent codec, \textbf{AudioVAE~V2}, that lifts the output sample rate to 48~kHz without lengthening the autoregressive sequence (Section~\ref{sec:method-audiovae}).
    \item A \textbf{refined backbone} with wider information pathways, a new isolated reference-audio input, and substantially scaled capacity (Section~\ref{sec:method-backbone}).
    \item A \textbf{unified sequence organization} that expresses basic TTS, voice design, reference cloning, controllable cloning, and continuation cloning as different input layouts over the same backbone (Section~\ref{sec:method-unified}).
\end{itemize}
The training strategy, data construction pipeline, and inference recipe are described in Sections~\ref{sec:method-training}--\ref{sec:method-inference}.

\subsection{AudioVAE V2}
\label{sec:method-audiovae}
The audio latent defines the representation on which the entire backbone operates. 
\modelname{} adopts \textbf{AudioVAE V2}, an asymmetric variational autoencoder whose encoder operates at 16~kHz and whose decoder reconstructs at 48~kHz. 
The asymmetric design serves two purposes simultaneously. 
On the decoder side, lifting the output rate to 48~kHz improves waveform fidelity to a high-quality regime without increasing the cost of the autoregressive generation loop.
On the encoder side, restricting the input rate to 16~kHz (i) enables seamless reuse of the large‑scale 16~kHz training corpus from the original VoxCPM, (ii) virtually eliminates latent mismatch across different source sample rates—thereby aligning the operational latent space of the backbone—and (iii) avoids the typical explosion in sequence length caused by a higher input rate.

Architecturally, AudioVAE~V2 follows the streaming-friendly causal-convolutional design of the original AudioVAE~\citep{zhou2025voxcpm} and modifies only the rate-related modules. 
The 16~kHz encoder uses a strided causal CNN stack with downsampling rates $[2, 5, 8, 8]$, yielding a 640$\times$ temporal reduction and producing 64-dimensional latent frames at 25~Hz. 
The 48~kHz decoder mirrors this structure with a deeper causal CNN stack and wider internal channels to support higher reconstruction bandwidth, with upsampling rates $[8, 6, 5, 2, 2, 2]$. 
Combined with the backbone patch size $P=4$, this yields a compact 6.25~Hz autoregressive sequence on the language-model side, sufficient for richer conditioning and longer contexts.
The decoder additionally accepts an optional target-sample-rate condition, allowing the same latent to be rendered at multiple actual output rates for downstream deployment.

\subsection{Backbone Refinements and Scaling}
\label{sec:method-backbone}
Extending VoxCPM into a multilingual and controllable foundation model imposes new requirements on the backbone, including higher conditioning bandwidth, support for arbitrary reference clips, and substantially increased capacity. 
We address these through three groups of changes: refining the internal architecture, adding an isolated reference-audio pathway, and scaling overall model capacity.

\subsubsection{Internal Architecture Refinements}
\label{sec:method-backbone-architecture}
\paragraph{Concatenation-projection fusion before RALM.} 
In VoxCPM, the FSQ-quantized semantic state $h^{\mathrm{FSQ}}_i$ and the Local Encoder embedding $E_i \in \mathbf{E}_{<i}$ were merged by element-wise summation before entering the RALM. 
\modelname{} replaces this with a learnable concatenation-projection,
\begin{equation}
    h^{\mathrm{res\_in}}_i = W_{\mathrm{fuse}} \,\bigl[\, h^{\mathrm{FSQ}}_i\;\Vert\; E_i \,\bigr],
\end{equation}
where $[\,\cdot\,\Vert\,\cdot\,]$ denotes channel-wise concatenation. 
This preserves richer information from both streams and allows the model to learn optimal combination weights.

\paragraph{Multi-token conditioning prefix in LocDiT.} 
In VoxCPM, the semantic state, the residual state, and the timestep embedding were summed into a single conditioning token. 
\modelname{} instead projects the three signals separately and feeds them as distinct prefix tokens to the LocDiT.
This avoids early information collapse and provides higher-bandwidth conditioning from the language model to the diffusion decoder.
The process is:
\[
    [\,\mu_{\mathrm{sem}},\; \mu_{\mathrm{res}},\; \mu_{t},\; z_{i-1}^{(1)},\,\ldots,\,z_{i-1}^{(P)},\; \tilde{z}_{i}^{(1)},\,\ldots,\,\tilde{z}_{i}^{(P)}\,],
\]
where $\mu_{\mathrm{sem}}$, $\mu_{\mathrm{res}}$, $\mu_{t}$ are the projections of $h^{\mathrm{FSQ}}_i$, $h^{\mathrm{residual}}_i$, and the diffusion timestep $t$, respectively, and $z^{(1\ldots P)}$ denotes the $P$ latent frames inside one patch (the previous clean patch and the current noisy patch). 
The LocDiT attends over this sequence with full attention and predicts the velocity field at the noisy-patch positions. 

\paragraph{Wider FSQ bottleneck.} 
We increase the FSQ bottleneck dimensionality from 256 to 512 to accommodate the larger model and broader linguistic coverage, while retaining the quantization granularity of 9 levels per dimension.

\paragraph{Removing positional encoding from RALM.} 
We retain Rotary Position Embeddings (RoPE)~\citep{su2024roformer} in the TSLM but remove them from the RALM following the NoPE design~\citep{kazemnejad2023impact}.
Since the RALM's role is primarily local acoustic refinement conditioned on the semantic skeleton, removing positional encodings reduces overfitting to training lengths and improves long-utterance stability.

\subsubsection{Reference Audio Pathway}
Beyond the continuation-style prompt inherited from VoxCPM, \modelname{} introduces an explicit \textit{reference-audio pathway}.
This pathway allows the insertion of a single reference audio clip from the target speaker as a voice-identity prefix, even without its transcript.
The reference clip is encoded by AudioVAE~V2 into latent patches and inserted as a delimited segment (\texttt{REF\_START}, \texttt{REF\_END}) at the beginning of the input sequence. 
Thanks to the causal nature of the TSLM and RALM, all subsequent positions can attend to this segment, providing robust speaker-identity information without requiring the reference to act as a temporal prefix of the target audio or to have an aligned transcript—in contrast to continuation-based cloning.
This decoupled design enables an optional voice cloning manner during inference without needing aligned text for the reference clip. It also lays the foundation for controllable cloning by effectively separating speaker identity from style control instructions.
The reference segment is excluded from the training loss and serves purely as conditioning context. 
Its integration with other input building blocks is detailed in Section~\ref{sec:method-unified}.

\subsubsection{Configuration Scaling}
Together with the above refinements, \modelname{} scales the backbone along depth, width, and context length.
Table~\ref{tab:config} summarizes the configuration relative to previous VoxCPM releases. 
A practically important update, first introduced in VoxCPM1.5 and retained here, is increasing the patch size from $P=2$ to $P=4$. 
This adjustment lowers the language-model-side token rate from 12.5~Hz to 6.25~Hz, reducing inference cost while improving long-form stability, consistent with recent trends in long-context speech modeling~\citep{peng2025vibevoice}.
Collectively, these changes enable the same hierarchical continuous-latent framework to scale from a bilingual zero-shot prototype to a large-scale multilingual and controllable foundation model, while preserving the compact token rate and streaming-friendly causal structure.

\begin{table}[h]
\centering
\caption{Configuration comparison across the VoxCPM family.}
\label{tab:config}
\resizebox{\linewidth}{!}{
\begin{tabular}{lccc}
\toprule
\textbf{Component} & \textbf{VoxCPM} & \textbf{VoxCPM1.5} & \textbf{\modelname{}} \\
\midrule
Backbone parameters & $\sim$0.6B & $\sim$0.8B & $\sim$2B \\
LocEnc & 4L, H=1024 & 8L, H=1024 & 12L, H=1024 \\
TSLM & MiniCPM-4-0.5B (24L, $H$=1024) & MiniCPM-4-0.5B (24L, $H$=1024) & MiniCPM-4-1B (28L, $H$=2048) \\
FSQ latent dim & 256 & 256 & 512 \\
RALM & 6L, H=1024 & 8L, H=1024 & 8L, H=2048 \\
LocDiT & 4L, H=1024 & 8L, H=1024 & 12L, H=1024 \\
Patch size $P$ & 2 & 4 & 4 \\
LM-side token rate & 12.5~Hz & 6.25~Hz & 6.25~Hz \\
Max sequence length & 4096 & 4096 & 8192 \\
Input sample rate & 16~kHz & 44.1~kHz & 16~kHz \\
Output sample rate & 16~kHz & 44.1~kHz & 48~kHz \\
\bottomrule
\end{tabular}
}
\end{table}

\subsection{Unified Sequence Organization}
\label{sec:method-unified}
\modelname{} supports five generation configurations through a single unified sequence organization rather than mode-specific modules. 
These configurations are built upon the same set of input building blocks, allowing the model to infer the intended behavior directly from the input layout. 
Although we refer to five configurations for completeness, they can be grouped into four primary capabilities: basic TTS, voice design, reference-based cloning (with or without additional style control), and continuation cloning.

The backbone processes two parallel tracks at each position—a text token and an audio latent—with a binary modality indicator determining the input embedding. 
An input sequence is assembled from three types of building blocks:
\begin{itemize}
    \item \textbf{text}: the synthesis transcription, optionally preceded by a natural-language description of the desired voice and/or style;
    \item \textbf{reference audio}: a delimited segment bracketed by \texttt{REF\_START}/\texttt{REF\_END} that supplies isolated voice-identity evidence;
    \item \textbf{target audio}: the segment the model is required to generate.
\end{itemize}
During training, only the target-audio segment contributes to the loss; preceding tokens serve as conditioning context. 
At inference, users may additionally supply a \textit{prompt audio} together with its transcript as observed context. 
This prompt is structurally treated as the initial prefix of the target audio segment used during training, from which the model continues autoregressively.
The five configurations differ only in how these building blocks are arranged, as summarized in Table~\ref{tab:modes}.

\begin{table}[h]
\centering
\small
\caption{Sequence layouts of the five generation configurations. ``$\rightarrow$'' separates the conditioning context from the target segment to be generated, and ``$|$'' separates building blocks within the conditioning context.}
\label{tab:modes}
\begin{tabular}{ll}
\toprule
\textbf{Mode} & \textbf{Sequence layout} \\
\midrule
Basic TTS               & $\langle$text$\rangle\;\rightarrow\;\langle$target audio$\rangle$ \\
Voice design            & $\langle$(voice description) text$\rangle\;\rightarrow\;\langle$target audio$\rangle$ \\
Reference cloning       & $\langle$reference audio$\rangle\;|\;\langle$text$\rangle\;\rightarrow\;\langle$target audio$\rangle$ \\
Controllable cloning    & $\langle$reference audio$\rangle\;|\;\langle$(style description) text$\rangle\;\rightarrow\;\langle$target audio$\rangle$ \\
Continuation cloning    & $\langle$prompt text $+$ target text$\rangle\;|\;\langle$prompt audio$\rangle\;\rightarrow\;\langle$target audio$\rangle$ \\
\bottomrule
\end{tabular}
\end{table}

Two aspects of this design are particularly noteworthy.
First, for voice design and controllable cloning, the natural-language description is simply concatenated with the synthesis text, allowing the same TSLM to handle both semantic content and control instructions without additional modules.
Second, continuation cloning benefits from a paired transcript for higher fidelity. At inference time, this layout can also be combined with an isolated reference segment to provide both temporal alignment and explicit speaker-identity evidence, yielding the ``Reference~$+$~Continuation'' recipe evaluated in Section~\ref{sec:exp-seedtts}.

\subsection{Training Strategy}
\label{sec:method-training}
\paragraph{Training objective.} 
We retain the two-term objective of VoxCPM: a patch-level conditional flow-matching loss on the target latent patches, and a binary stop-prediction loss on the TSLM-FSQ hidden states. 
Both losses are masked to the target-audio segment only. 
To enable classifier-free guidance at inference (Section~\ref{sec:method-inference}), we randomly drop the LM-side conditioning of the LocDiT with probability $10\%$ during training. 
The optimizer is AdamW with cosine learning-rate decay and linear warmup.

\paragraph{Three-stage progressive curriculum.} 
To prevent destabilizing the base synthesis quality when incorporating all target capabilities simultaneously, we adopt a three-stage progressive curriculum.
The loss formulation remains fixed across all stages, while we vary only the data composition, mixing ratio, and context length:
\begin{enumerate}
    \item \textbf{Multilingual TTS pretraining.} 
    The backbone is trained on large-scale multilingual <transcription, audio> pairs for basic TTS and continuation cloning. Audio segments are limited to 60~s and the maximum LM sequence length is set to 4096 to ensure stable and fast optimization.
    This stage establishes solid pronunciation and prosody across all 30 target languages.
    \item \textbf{Joint TTS and controllable TTS pretraining.} 
    Building upon stage 1, we retain a large proportion of plain TTS data to preserve base synthesis quality, while gradually introducing controllable data at an increasing ratio.
    This includes (i) speech annotated with natural-language voice and style descriptions to supervise voice design, and (ii) <reference audio, transcription, target audio> triplets to train both reference-based and controllable cloning. 
    We extend the maximum sequence length to 8192 and audio duration to up to 3 minutes. 
    Stages 1 and 2 together account for the majority of training compute.
    \item \textbf{High-quality annealing SFT.} 
    The final stage uses a curated high-quality subset with more expressive speech and precisely annotated controllable data.
    Controllable samples occupy a significantly larger proportion than in Stage 2, including a higher ratio of reference-audio-based controllable cloning examples.
    We maintain the 8192-token context, utilize diverse samples ranging from 2~sec to 5~min, adopt a balanced language-level sampling ratio, and apply learning-rate annealing to further refine performance.
\end{enumerate}

\subsection{Data Construction and Annotation for Controllability}
\label{sec:method-data}
The total training corpus comprises over 2 million hours of multilingual speech, with Chinese and English forming the majority. 
The remaining 28 languages range from roughly 1\,K to 50\,K hours each, depending on data availability and annotation quality.
The base TTS data follows a standard preparation pipeline: source separation, voice activity detection, ASR-based transcript alignment, and quality filtering.

For controllable generation, we combine tens of thousands of hours of open-source expressive speech with several thousand hours of internally curated and annotated data. 
The open-source portion provides broad coverage of emotions, speaking styles, and speakers, while the internal portion emphasizes higher annotation precision and richer natural-language descriptions. 
The remainder of this section details how the internal subset is constructed and annotated.

\paragraph{Selecting expressive audio worth annotating.} 
Public controllable corpora often contain acoustically flat utterances, capping the upper bound of controllability. 
To avoid this, we first collect speech across diverse expressive scenarios, and then pre-screen large unlabeled corpora using lightweight emotion classifiers, retaining only sufficiently expressive samples for annotation.

\paragraph{Multi-dimensional natural-language annotation.} 
We annotate selected expressive utterances along two axes that mirror the target capabilities: 
voice design attributes (e.g., age, gender, accent, vocal texture, scenario) and style control attributes (e.g., emotion, speaking rate, pitch, energy, emphasis).
Annotations are generated using general-purpose audio understanding models (e.g., Step-Audio R1~\citep{tian2025step} and Gemini 2.5 Pro) which produce free-form natural-language descriptions at varying granularities, and verified with some dedicated expert classifiers in terms of gender, age and emotion.
The resulting descriptions are used directly as text prefixes, requiring no additional embedding modules.

\paragraph{Mining same-speaker references for cloning.} 
Reference-based cloning requires a reference clip that shares the speaker with the target utterance. 
We harvest reference clips from the same recording session by computing speaker-embedding cosine similarity and retaining clips with a similarity score above 0.7. We additionally exclude clips that directly precede the target utterance.
This ensures they do not directly precede the target utterance. 
Note that even with this threshold, the selected reference clips may still differ in fine acoustic details from the target, so reference-based cloning naturally achieves lower similarity than continuation-based cloning. 
This pool supports both reference cloning and controllable cloning.

\paragraph{Decoupling style from content via cloned synthesis.} 
A common challenge with naturally annotated expressive speech is the strong correlation between prosodic style and textual content (e.g., cheerful styles often co-occur with positive sentences).  
Training directly on such data risks the model recovering style from the text rather than from the control prompt, thereby weakening controllability.
To address this, we use the model itself to generate content-decoupled examples: starting from an annotated utterance, we clone its voice and style onto a semantically unrelated transcript, while preserving the original natural-language description as the control prompt. The resulting <description, another text, audio> pairs, whose content no longer leaks style cues, are mixed back into training. 
This procedure also helps extend controllable training to long-tail languages with limited native expressive data.
To minimize potential artifacts from self-synthesized speech, we inject this data primarily in stage 2 and restrict the stage 3 annealing mixture to natively recorded high-quality speech.

\subsection{Inference}
\label{sec:method-inference}

At inference time, \modelname{} generates speech autoregressively, one latent patch at a time. 
We adopt three techniques to balance speed and quality.

\paragraph{Classifier-free guidance (CFG).}
The LM-side conditioning to the LocDiT is randomly dropped during training, enabling both conditional and unconditional predictions. 
At each denoising step, we evaluate the LocDiT twice and linearly combine the velocity fields as $ \hat{v} = v_{\mathrm{uncond}} + \alpha\,(v_{\mathrm{cond}} - v_{\mathrm{uncond}})$, where $\alpha=2.0$ by default. 
We find $\alpha \in [1.5,\,3.0]$ to be a practical range.

\paragraph{Sway sampling and CFG-Zero*.}
We apply sway sampling~\citep{chen2024f5} to allocate more solver steps to high-noise regimes and CFG-Zero*~\citep{fan2025cfgzerostar} to reduce early-step artifacts. 
Both techniques are enabled by default and introduce no additional learnable parameters.

\paragraph{Streaming.}
The causal structure of the TSLM and RALM, combined with the patch-local design of LocEnc and LocDiT (each operating over a single patch with full intra-patch attention), enables efficient patch/chunk-based streaming. 
Each generated latent patch is immediately decoded by a stateful AudioVAE~V2 decoder.
For continuation mode, the last few prompt patches are retained as the decoder's initial context to ensure smooth transitions.

\section{Experiments and Results}
\label{sec:exp}
We evaluate \modelname{} on zero-shot voice cloning, multilingual synthesis, natural-language controllability, reconstruction quality, and inference efficiency.
Experiments are conducted on public benchmarks and internal test sets. 
We report zero-shot cloning performance in Section~\ref{sec:exp-seedtts}, multilingual results in Section~\ref{sec:exp-multilingual}, controllable generation in Section~\ref{sec:exp-control}, AudioVAE~V2 reconstruction quality in Section~\ref{sec:exp-audiovae}, inference efficiency in Section~\ref{sec:exp-efficiency}, and subjective listening tests in Section~\ref{sec:exp-subjective}.

\subsection{Experimental Setup}
\label{sec:exp-setup}

\paragraph{Benchmarks.}
For zero-shot and multilingual synthesis, we use three public benchmarks: (i)~\textbf{Seed-TTS-Eval}\footnote{\url{https://github.com/BytedanceSpeech/seed-tts-eval}~\citep{anastassiou2024seed}}, a Chinese-English voice cloning benchmark with two standard test sets and a more challenging \textit{hard} subset; (ii)~\textbf{CV3-Eval}\footnote{\url{https://github.com/FunAudioLLM/CV3-Eval}}~\citep{du2025cosyvoice3}, an in-the-wild multilingual zero-shot cloning benchmark covering nine languages with additional hard subsets for Chinese and English, which features more diverse expressive styles and audio qualities in the reference clips; (iii)~\textbf{MiniMax-MLS-Test}\footnote{\url{https://huggingface.co/datasets/MiniMaxAI/TTS-Multilingual-Test-Set}}~\citep{zhang2025minimax}, another multilingual zero-shot voice cloning benchmark spanning 24 languages. 
For natural-language-guided controllable generation, we adopt \textbf{InstructTTSEval}\footnote{\url{https://github.com/KexinHUANG19/InstructTTSEval}~\citep{huang2025instructttseval}}, which decomposes instruction following into three subtasks of increasing abstraction: \textbf{APS} (acoustic-parameter specification), \textbf{DSD} (descriptive-style directive), and \textbf{RP} (role-play). 
To better evaluate languages not fully covered by public benchmarks, we constructed an \textbf{Internal 30-Language Benchmark} consisting of 500 utterances per language.
The reference audio clips for cloning evaluation were collected from CommonVoice\footnote{\url{https://commonvoice.mozilla.org/}} and Fleurs\footnote{\url{https://huggingface.co/datasets/google/fleurs}}. 
For public benchmarks, intelligibility is evaluated with the benchmark-standard or previously reported ASR setup when applicable, and with Whisper-large-v3 on MiniMax-MLS-Test for consistency with prior comparisons. 
For our internal 30-language benchmark, we use the Gemini 3.1 Flash Lite API for ASR transcription, as Whisper-large-v3 shows limited accuracy on several low-resource languages.

\paragraph{Comparison Systems.}
We compare \modelname{} against a diverse set of representative systems, including strong open-source baselines and recent state-of-the-art models  (CosyVoice family, MaskGCT, Spark-TTS, FireRedTTS series, F5-TTS, Qwen3-TTS, IndexTTS2, VibeVoice, HiggsAudio-v2, MOSS-TTS, Fish Audio S2, LongCat-Audio-DiT, as well as closed-source systems such as MegaTTS3, DiTAR, Seed-TTS, MiniMax-Speech, ElevenLabs, and Hume).
Earlier VoxCPM versions (VoxCPM and VoxCPM1.5) are included as internal references.

\paragraph{Metrics.}
We adopt widely used objective metrics following community standards.
For intelligibility, we report \textbf{WER} (Word Error Rate) for English and most European languages, and \textbf{CER} (Character Error Rate) for Chinese and other character-based languages. 
Speaker similarity is measured by \textbf{SIM}, the cosine similarity of speaker embeddings extracted from a pretrained speaker verification model. 
For controllable generation on InstructTTSEval, we report the official instruction-following accuracy for each subtask (APS, DSD, and RP).

\subsection{Zero-Shot Voice Cloning on Seed-TTS-Eval}
\label{sec:exp-seedtts}

\begin{table}[t]
\centering
\small
\caption{Zero-shot voice cloning on Seed-TTS-Eval. WER (English) / CER (Chinese, Hard) reported in \%; SIM in \%. Bold marks the strongest open-source result per column; italic marks the strongest closed-source result per column. ``--'' indicates the result is not reported in the source publication or is unavailable.} 
\label{tab:seedtts}
\setlength{\tabcolsep}{4pt}
\begin{tabular}{lccccccccc}
\toprule
\multirow{2}{*}{\textbf{Model}} & \multirow{2}{*}{\textbf{Params}} & \multirow{2}{*}{\textbf{OS}} & \multicolumn{2}{c}{\textbf{test-EN}} & \multicolumn{2}{c}{\textbf{test-ZH}} & \multicolumn{2}{c}{\textbf{test-ZH-Hard}} \\
\cmidrule(lr){4-5} \cmidrule(lr){6-7} \cmidrule(lr){8-9}
 &  &  & WER$\downarrow$ & SIM$\uparrow$ & CER$\downarrow$ & SIM$\uparrow$ & CER$\downarrow$ & SIM$\uparrow$ \\
\midrule
\multicolumn{9}{l}{\textit{Closed-source}}\\
MegaTTS3~\citep{jiang2025megatts3}                 & 0.5B  & \ding{55} & 2.79 & \textit{77.1} & 1.52 & 79.0 & --   & --   \\
DiTAR~\citep{jia2025ditar}                    & 0.6B  & \ding{55} & 1.69 & 73.5 & 1.02 & 75.3 & --   & --   \\
CosyVoice 3~\citep{du2025cosyvoice3}               & 1.5B  & \ding{55} & 2.22 & 72.0 & 1.12 & 78.1 & 5.83 & 75.8 \\
Seed-TTS~\citep{anastassiou2024seed}                 & --    & \ding{55} & 2.25 & 76.2 & 1.12 & 79.6 & 7.59 & 77.6 \\
MiniMax-Speech~\citep{zhang2025minimax}           & --    & \ding{55} & 1.65 & 69.2 & \textit{0.83} & 78.3 & --   & --   \\
CosyVoice3.5            & --  & \ding{55} & \textit{1.57} & 73.8 & 0.87 & \textit{79.7} & \textit{5.71} & \textit{78.6} \\
\midrule
\multicolumn{9}{l}{\textit{Open-source}}\\
F5-TTS~\citep{chen2024f5}                   & 0.3B  & \ding{51} & 2.00 & 67.0 & 1.53 & 76.0 & 8.67 & 71.3 \\
MaskGCT~\citep{wangmaskgct}                  & 1B    & \ding{51} & 2.62 & 71.7 & 2.27 & 77.4 & --   & --   \\
CosyVoice~\citep{du2024cosyvoice1}                & 0.3B  & \ding{51} & 4.29 & 60.9 & 3.63 & 72.3 & 11.75 & 70.9 \\
CosyVoice 2~\citep{du2024cosyvoice2}              & 0.5B  & \ding{51} & 3.09 & 65.9 & 1.38 & 75.7 & 6.83 & 72.4 \\
CosyVoice 3~\citep{du2025cosyvoice3}              & 0.5B  & \ding{51} & 2.02 & 71.8 & 1.16 & 78.0 & 6.08 & 75.8 \\
Spark-TTS~\citep{wang2025spark}                 & 0.5B  & \ding{51} & 3.14 & 57.3 & 1.54 & 66.0 & --   & --   \\
FireRedTTS~\citep{guo2024fireredtts}               & 0.5B  & \ding{51} & 3.82 & 46.0 & 1.51 & 63.5 & 17.45 & 62.1 \\
FireRedTTS-2~\citep{xie2025fireredtts}             & 1.5B  & \ding{51} & 1.95 & 66.5 & 1.14 & 73.6 & --   & --   \\
Qwen2.5-Omni~\citep{xu2025qwen2}             & 7B    & \ding{51} & 2.72 & 63.2 & 1.70 & 75.2 & 7.97 & 74.7 \\
Qwen3-Omni~\citep{xu2025qwen3}               & 30B-A3B & \ding{51} & 1.39 & --   & 1.07 & --   & --   & --   \\
OpenAudio-s1-mini~\citep{openaudios1}        & 0.5B  & \ding{51} & 1.94 & 55.0 & 1.18 & 68.5 & 23.37 & 64.3 \\
IndexTTS2~\citep{zhou2025indextts2}                & 1.5B  & \ding{51} & 2.23 & 70.6 & 1.03 & 76.5 & 7.12 & 75.5 \\
VibeVoice~\citep{peng2025vibevoice}                & 1.5B  & \ding{51} & 3.04 & 68.9 & 1.16 & 74.4 & --   & --   \\
HiggsAudio-v2~\citep{bosonai2025higgsaudio}            & 3B    & \ding{51} & 2.44 & 67.7 & 1.50 & 74.0 & 55.07 & 65.6 \\
ZipVoice~\citep{zhu2025zipvoice}                 &  0.1B   & \ding{51} & 1.64 & 66.8 & 1.40 & 75.1 & -- & -- \\
MOSS-TTS~\citep{gong2026moss}                 & 8B    & \ding{51} & 1.85 & 73.4 & 1.20 & 78.8 & --   & --   \\
Qwen3-TTS~\citep{hu2026qwen3}                & 1.7B  & \ding{51} & 1.23 & 71.7 & 1.22 & 77.0 & 6.76 & 74.8 \\
Fish Audio S2~\citep{liao2026fish}             & 4B    & \ding{51} & \textbf{0.99} & --   & \textbf{0.54} & --   & \textbf{5.99} & --   \\
OmniVoice~\citep{zhu2026omnivoice}                 &  0.8B   & \ding{51} & 1.60 & 74.1 & 0.84 & 77.7 & -- & -- \\
LongCat-Audio-DiT~\citep{xin2026longcat}        & 3.5B  & \ding{51} & 1.50 & \textbf{78.6} & 1.09 & \textbf{81.8} & 6.04 & \textbf{79.7} \\
\midrule
VoxCPM        & 0.6B  & \ding{51} & 1.85 & 72.9 & 0.93 & 77.2 & 8.87 & 73.0 \\
VoxCPM1.5               & 0.8B  & \ding{51} & 2.12 & 71.4 & 1.18 & 77.0 & 7.74 & 73.1 \\
\textbf{\modelname{}}    & 2B    & \ding{51} & 1.84 & 75.3 & 0.97 & 79.5 & 8.13 & 75.3 \\
\bottomrule
\end{tabular}
\end{table}

Table~\ref{tab:seedtts} presents zero-shot voice cloning results on Seed-TTS-Eval. 
At 2B parameters, \modelname{} achieves competitive performance among both open-source and closed-source systems, attaining \textbf{1.84/75.3} (WER/SIM) on test-EN, \textbf{0.97/79.5} on test-ZH, and \textbf{8.13/75.3} on the challenging test-ZH-Hard subset.
As a hierarchical continuous-latent model, \modelname{} demonstrates strong speaker similarity while maintaining good intelligibility.
It outperforms most token-based autoregressive systems in similarity and achieving comparable or better WER/CER than many non-autoregressive models. 
Subjective listening tests further confirm that its cloning performance is superior in terms of naturalness and prosody.

Despite the significant increase in capabilities and language coverage, the architectural refinements and parameter scaling allow \modelname{} to maintain the strong zero-shot cloning ability of previous VoxCPM versions while further improving speaker similarity.

\paragraph{Inference recipes.}
Section~\ref{sec:method-unified} introduces three inference recipes for conditioning on a reference utterance: (i)~\textit{continuation only}, treating the reference as an audio prefix paired with its transcript, identical to the operating mode of VoxCPM and VoxCPM1.5; 
(ii)~\textit{reference only}, placing the reference in the isolated reference-audio segment, without using its transcript; 
and (iii)~\textit{reference~+~continuation}, supplying the same reference clip in both pathways. 
Table~\ref{tab:infer_recipe} compares their effectiveness on Seed-TTS-Eval.
Three main observations can be drawn. 
First, combining both pathways (reference + continuation) consistently yields the best overall performance, achieving the highest SIM across all subsets. 
The two mechanisms are complementary: the continuation prefix provides temporal prosodic alignment, while the isolated reference supplies robust speaker identity. 
Second, the reference-only recipe achieves the best intelligibility on hard Chinese subsets. 
Without a temporal audio prefix, the model gains more flexibility in prosody selection for complex content, at a modest cost in similarity. 
Third, continuation-only remains a strong baseline, particularly for easier cases. 
Unless otherwise specified, all other VoxCPM2 zero-shot cloning results in this paper use the reference + continuation recipe.

\begin{table}[t]
\centering
\small
\caption{Effect of different inference recipes on \modelname{} zero-shot voice cloning performance (Seed-TTS-Eval).}
\label{tab:infer_recipe}
\setlength{\tabcolsep}{5pt}
\begin{tabular}{llcccccc}
\toprule
\multirow{2}{*}{\textbf{Inference recipe}} & \multirow{2}{*}{\textbf{Sequence layout (Sec.~3.4)}} & \multicolumn{2}{c}{\textbf{test-EN}} & \multicolumn{2}{c}{\textbf{test-ZH}} & \multicolumn{2}{c}{\textbf{test-ZH-Hard}} \\
\cmidrule(lr){3-4} \cmidrule(lr){5-6} \cmidrule(lr){7-8}
 & & WER$\downarrow$ & SIM$\uparrow$ & CER$\downarrow$ & SIM$\uparrow$ & CER$\downarrow$ & SIM$\uparrow$ \\
\midrule
Continuation only & $\langle\text{text}\,|\,\text{prompt}\rangle\!\rightarrow\!\langle\text{target}\rangle$ & 1.01 & 77.7 & 1.97 & 72.6 & 8.16 & 72.4 \\
Reference only & $\langle\text{ref}\rangle\,|\,\langle\text{text}\rangle\!\rightarrow\!\langle\text{target}\rangle$ & 1.10 & 75.3 & \textbf{1.81} & 67.0 & \textbf{6.85} & 70.0 \\
\textbf{Reference + Continuation} & $\langle\text{ref}\rangle\,|\,\langle\text{text}\,|\,\text{prompt}\rangle\!\rightarrow\!\langle\text{target}\rangle$ & \textbf{0.99} & \textbf{79.5} & 1.94 & \textbf{75.2} & 7.44 & \textbf{74.9} \\
\bottomrule
\end{tabular}
\end{table}

\subsection{Multilingual Capability}
\label{sec:exp-multilingual}
We evaluate the multilingual capability of \modelname{}, focusing not only on intelligibility but also on whether strong speaker preservation can be maintained across a broad range of languages. 
Results are reported on three benchmarks: CV3-Eval (Table~\ref{tab:cv3eval}), MiniMax-MLS-Test (Tables~\ref{tab:mls_wer} and~\ref{tab:mls_sim}), and an internal 30-language benchmark(Table~\ref{tab:internal_30lang}).

\paragraph{CV3-Eval: In-the-wild Multilingual Cloning.}
CV3-Eval is a challenging in-the-wild benchmark with diverse expressive styles and audio conditions. 
\modelname{} demonstrates strong stability, achieving competitive intelligibility across all nine languages, particularly on the hard subsets (hard-zh: 8.55, hard-en: 8.48). 
While Fish Audio S2 attains lower WER on many languages thanks to its larger scale and additional RL post-training, \modelname{} remains highly competitive given its smaller 2B size and fully end-to-end continuous-latent design without discrete speech tokenizers.

\begin{table}[t]
\centering
\small
\caption{CV3-Eval multilingual zero-shot cloning. WER/CER (\%, lower is better). Bold marks the best result per column. ``--'' indicates the result is not reported in the source publication.}
\label{tab:cv3eval}
\setlength{\tabcolsep}{3pt}
\begin{tabular}{lccccccccccc}
\toprule
\textbf{Model} & \textbf{zh} & \textbf{en} & \textbf{hard-zh} & \textbf{hard-en} & \textbf{ja} & \textbf{ko} & \textbf{de} & \textbf{es} & \textbf{fr} & \textbf{it} & \textbf{ru} \\
\midrule
CosyVoice 2          & 4.08 & 6.32 & 12.58 & 11.96 & 9.13 & 19.70 & --   & --   & --    & --    & --   \\
CosyVoice 3-1.5B     & 3.91 & 4.99 & 9.77  & 10.55 & 7.57 & 5.69  & 6.43 & 4.47 & 11.80 & 10.50 & 6.64 \\
Fish Audio S2         & \textbf{2.65} & \textbf{2.43} & 9.10 & \textbf{4.40} & \textbf{3.96} & \textbf{2.76} & \textbf{2.22} & \textbf{2.00} & \textbf{6.26} & \textbf{2.04} & \textbf{2.78} \\
\textbf{\modelname{}} & 3.65 & 5.00 & \textbf{8.55} & 8.48 & 5.96 & 5.69 & 4.77 & 3.80 & 9.85 & 4.25 & 5.21 \\
\bottomrule
\end{tabular}
\end{table}

\paragraph{MiniMax-MLS-Test: 24-language Intelligibility and Speaker Similarity.}
On speaker similarity (Table~\ref{tab:mls_sim}), \modelname{} achieves the highest SIM on 22 out of 24 languages, demonstrating strong and consistent voice preservation across diverse language families. 
Even on Czech and Romanian—languages not explicitly seen during training---\modelname{} shows promising speaker preservation, falling within only 1.5–2 SIM points of the best system. 
This advantage is largely attributed to the hierarchical continuous-latent modeling paradigm, which enables richer speaker representation compared to discrete-token approaches. Additionally, large-scale multilingual pretraining appears to endow the model with a certain degree of emergent zero-shot synthesis capability on languages within similar families.
On intelligibility (Table~\ref{tab:mls_wer}), \modelname{} delivers strong results on most languages, particularly Chinese (1.14), Dutch (0.91), Finnish (2.63), German (0.68), and Turkish (0.82). 
The main weaknesses appear in Arabic and Hindi. 
These languages are included in the training set but with relatively limited data volume. 
Moreover, since we use Whisper-large-v3 for ASR evaluation on this benchmark, part of the higher WER may stem from the recognizer’s limited accuracy on these languages rather than purely from synthesis quality. 
For completely unseen languages such as Czech and Romanian, the model still produces partially intelligible speech, indicating partial zero-shot transfer and suggesting room for rapid adaptation through few-shot fine-tuning

\begin{table}[t]
\centering
\small
\caption{Intelligibility on MiniMax-MLS-Test (24 languages, WER\,\%, lower is better). Bold marks the best result per language; ``--'' indicates the system does not report a result for the language.}
\label{tab:mls_wer}
\setlength{\tabcolsep}{4pt}
\begin{tabular}{lccccc}
\toprule
\textbf{Language} & \textbf{Minimax} & \textbf{ElevenLabs} & \textbf{Qwen3-TTS} & \textbf{Fish Audio S2} & \textbf{\modelname{}} \\
\midrule
Arabic        & \textbf{1.67} & 1.67  & --   & 3.50  & 13.05 \\
Cantonese     & 34.11 & 51.51 & --   & \textbf{30.67} & 38.58 \\
Chinese       & 2.25  & 16.03 & 0.93 & \textbf{0.73}  & 1.14  \\
Czech         & 3.88  & \textbf{2.11} & --   & 2.84  & 24.13 \\
Dutch         & 1.14  & \textbf{0.80} & --   & 0.99  & 0.91  \\
English       & 2.16  & 2.34  & \textbf{0.93} & 1.62  & 2.29  \\
Finnish       & 4.67  & 2.96  & --   & 3.33  & \textbf{2.63} \\
French        & 4.10  & 5.22  & \textbf{2.86} & 3.05  & 4.53  \\
German        & 1.91  & 0.57  & 1.24 & \textbf{0.55}  & 0.68  \\
Greek         & 2.02  & \textbf{0.99} & --   & 5.74  & 2.84  \\
Hindi         & 6.96  & \textbf{5.83} & --   & 14.64 & 19.70 \\
Indonesian    & 1.24  & \textbf{1.06} & --   & 1.46  & 1.08  \\
Italian       & 1.54  & 1.74  & \textbf{0.95} & 1.27  & 1.56  \\
Japanese      & 3.52  & 10.65 & 3.82 & \textbf{2.76}  & 4.63  \\
Korean        & 1.75  & 1.87  & 1.76 & \textbf{1.18}  & 1.96  \\
Polish        & 1.42  & \textbf{0.77} & --   & 1.26  & 1.14  \\
Portuguese    & 1.88  & 1.33  & 1.53 & \textbf{1.14}  & 1.94  \\
Romanian      & 2.88  & \textbf{1.35} & --   & 10.74 & 21.58 \\
Russian       & 4.28  & 3.88  & 3.21 & \textbf{2.40}  & 3.63  \\
Spanish       & 1.03  & 1.08  & 1.13 & \textbf{0.91}  & 1.44  \\
Thai          & \textbf{2.70} & 73.94 & --   & 4.23  & 2.96  \\
Turkish       & 1.52  & \textbf{0.70} & --   & 0.87  & 0.82  \\
Ukrainian     & 1.08  & \textbf{1.00} & --   & 2.30  & 6.32  \\
Vietnamese    & \textbf{0.88} & 73.42 & --   & 7.41  & 3.31  \\
\bottomrule
\end{tabular}
\end{table}

\begin{table}[t]
\centering
\small
\caption{Speaker similarity on MiniMax-MLS-Test (24 languages, SIM\,\%, higher is better). Bold marks the best result per language.}
\label{tab:mls_sim}
\setlength{\tabcolsep}{4pt}
\begin{tabular}{lccccc}
\toprule
\textbf{Language} & \textbf{Minimax} & \textbf{ElevenLabs} & \textbf{Qwen3-TTS} & \textbf{Fish Audio S2} & \textbf{\modelname{}} \\
\midrule
Arabic        & 73.6 & 70.6 & --   & 75.0 & \textbf{79.1} \\
Cantonese     & 77.8 & 67.0 & --   & 80.5 & \textbf{83.5} \\
Chinese       & 78.0 & 67.7 & 79.9 & 81.6 & \textbf{82.5} \\
Czech         & 79.6 & 68.5 & --   & \textbf{79.8} & 78.3 \\
Dutch         & 73.8 & 68.0 & --   & 73.0 & \textbf{80.8} \\
English       & 75.6 & 61.3 & 77.5 & 79.7 & \textbf{85.4} \\
Finnish       & 83.5 & 75.9 & --   & 81.9 & \textbf{89.0} \\
French        & 62.8 & 53.5 & 62.8 & 69.8 & \textbf{73.5} \\
German        & 73.3 & 61.4 & 77.5 & 76.7 & \textbf{80.3} \\
Greek         & 82.6 & 73.3 & --   & 79.5 & \textbf{86.0} \\
Hindi         & 81.8 & 73.0 & --   & 82.1 & \textbf{85.6} \\
Indonesian    & 72.9 & 66.0 & --   & 76.3 & \textbf{80.0} \\
Italian       & 69.9 & 57.9 & 81.7 & 74.7 & \textbf{78.0} \\
Japanese      & 77.6 & 73.8 & 78.8 & 79.6 & \textbf{82.8} \\
Korean        & 77.6 & 70.0 & 79.9 & 81.7 & \textbf{83.3} \\
Polish        & 80.2 & 72.9 & --   & 81.9 & \textbf{88.4} \\
Portuguese    & 80.5 & 71.1 & 81.7 & 78.1 & \textbf{83.7} \\
Romanian      & \textbf{80.9} & 69.9 & --   & 73.3 & 79.7 \\
Russian       & 76.1 & 67.6 & 79.2 & 79.0 & \textbf{81.1} \\
Spanish       & 76.2 & 61.5 & 81.4 & 77.6 & \textbf{83.1} \\
Thai          & 80.0 & 58.8 & --   & 78.6 & \textbf{84.0} \\
Turkish       & 77.9 & 59.6 & --   & 83.5 & \textbf{87.1} \\
Ukrainian     & 73.0 & 64.7 & --   & 74.7 & \textbf{79.8} \\
Vietnamese    & 74.3 & 36.9 & --   & 74.0 & \textbf{80.6} \\
\bottomrule
\end{tabular}
\end{table}

\paragraph{Internal 30-Language Benchmark.}
To better evaluate performance on languages that are not fully covered by existing public benchmarks, we constructed an internal 30-language test set mentioned above. On this benchmark, \modelname{} achieves an average WER/CER of 1.68\% across all 30 languages, with error rates below 3\% on 28 languages and below 1\% on six languages. 
The model shows particularly strong results on several Southeast Asian and low-resource languages such as Khmer, Lao, Burmese, and Thai.
Notably, performance on Arabic and Hindi improved substantially compared to evaluations using Whisper-large-v3 on Minimax-MLS-Test. 
Since these results were obtained using the Gemini 3.1 Flash Lite API for ASR transcription, this further supports our earlier observation regarding the limitations of Whisper on certain languages. 
Overall, these findings indicate that \modelname{} provides a competitive and balanced multilingual TTS option within a single unified hierarchical continuous-latent model.

\begin{table}[t]
\centering
\small
\caption{Internal 30-language intelligibility benchmark (500 utterances per language; ASR by Gemini~3.1 Flash Lite). WER\,\% for word-segmented languages; CER\,\% for logographic or non-segmented scripts. }
\label{tab:internal_30lang}
\setlength{\tabcolsep}{4pt}
\begin{tabular}{llcc|llcc}
\toprule
\textbf{Lang.} & \textbf{Metric} & \textbf{\modelname{}} & \textbf{Fish Audio S2} &
\textbf{Lang.} & \textbf{Metric} & \textbf{\modelname{}} & \textbf{Fish Audio S2} \\
\midrule
ar (Arabic)     & CER & 1.23 & 0.30 & lo (Lao)      & CER & 1.90 & 87.40  \\
da (Danish)     & WER & 2.70 & 3.52 & ms (Malay)    & WER & 1.75 & 1.41 \\
de (German)     & WER & 0.96 & 0.64 & my (Burmese)  & CER & 1.42 & 85.27 \\
el (Greek)      & WER & 3.17 & 4.61 & nl (Dutch)    & WER & 1.25 & 1.68 \\
en (English)    & WER & 0.42 & 1.03 & no (Norwegian)& WER & 2.49 & 3.76 \\
es (Spanish)    & WER & 1.33 & 0.64 & pl (Polish)   & WER & 1.90 & 1.65 \\
fi (Finnish)    & WER & 2.24 & 2.80 & pt (Portuguese)& WER & 1.48 & 1.49 \\
fr (French)     & WER & 2.16 & 2.34 & ru (Russian)  & WER & 0.90 & 0.86 \\
he (Hebrew)     & CER & 2.98 & 15.27 & sv (Swedish)  & WER & 2.22 & 2.63 \\
hi (Hindi)      & CER & 0.79 & 0.91 & sw (Swahili)  & CER & 1.07 & 2.02 \\
id (Indonesian) & WER & 1.36 & 1.68 & th (Thai)     & CER & 0.94 & 1.92 \\
it (Italian)    & WER & 1.65 & 1.08 & tl (Tagalog)  & WER & 2.63 & 4.00 \\
ja (Japanese)   & CER & 2.40 & 1.82 & tr (Turkish)  & WER & 1.65 & 1.65 \\
km (Khmer)      & CER & 2.05 & 75.15 & vi (Vietnamese)& WER & 1.56 & 5.56 \\
ko (Korean)   & CER & 0.95 & 0.29 & zh (Chinese)  & CER & 0.92 & 1.02 \\
\midrule
\multicolumn{6}{r}{\textbf{Average across 30 languages}} & \textbf{1.68} & -- \\
\bottomrule
\end{tabular}
\end{table}

\subsection{Voice Design and Controllable Generation}
\label{sec:exp-control}
Beyond zero-shot voice cloning, \modelname{} additionally supports \textit{voice design}, which synthesizes a new voice from a natural-language description without any reference audio, and \textit{controllable cloning}, which clones a speaker from a reference clip while following style instructions. 
Both capabilities are implemented through the same unified sequence organization (Section~\ref{sec:method-unified}), without any dedicated control heads or style encoders.

InstructTTSEval provides a comprehensive assessment of instruction-following ability across three subtasks of increasing abstraction.
Table~\ref{tab:instructtts} summarizes the results.
On this benchmark, \modelname{} achieves strong instruction-following performance. 
On the English subset, it attains the best overall scores with \textbf{84.2 / 83.2 / 71.4} on APS, DSD, and RP respectively, outperforming all compared systems. 
On the Chinese subset, it matches the top APS score (85.2, tied with Qwen3-TTS) and remains competitive on DSD (71.5) and RP (60.8), though it trails slightly on the more abstract tasks.
The gap on the more abstract Chinese tasks may be attributed to current limitations in annotation diversity for higher-level stylistic and persona descriptions.

The controllable generation capability of \modelname{} extends well beyond the InstructTTSEval benchmark. 
The model supports a wide variety of natural-language descriptions for voice design and style control, and can generate speech in all 30 supported languages as well as 9 Chinese dialects. More diverse examples and interactive demonstrations are available on the project demo page. Additionally, thanks to the inclusion of song-style annotated data during training, \modelname{} also exhibits preliminary singing voice generation ability, although singing quality remains an area for further improvement.

\begin{table}[t]
\centering
\small
\caption{Instruction-following on InstructTTSEval (\%, higher is better). APS = acoustic-parameter specification, DSD = descriptive-style directive, RP = role-play.}
\label{tab:instructtts}
\setlength{\tabcolsep}{6pt}
\begin{tabular}{lcccccc}
\toprule
\multirow{2}{*}{\textbf{Model}} & \multicolumn{3}{c}{\textbf{InstructTTSEval-ZH}} & \multicolumn{3}{c}{\textbf{InstructTTSEval-EN}} \\
\cmidrule(lr){2-4} \cmidrule(lr){5-7}
 & APS$\uparrow$ & DSD$\uparrow$ & RP$\uparrow$ & APS$\uparrow$ & DSD$\uparrow$ & RP$\uparrow$ \\
\midrule
Hume                 & --   & --   & --   & 83.0 & 75.3 & 54.3 \\
GPT-4o-mini-TTS      & 54.9 & 52.3 & 46.0   & 76.4 & 74.3 & 54.8 \\
Gemini-TTS-Pro      & 89.0 & 90.1 & 75.5 & 87.6	 & 86.0	 & 67.2	 \\
\midrule
PromptTTS~\citep{guo2023prompttts}                     & --   & --   & --   & 64.3 & 47.2 & 31.4 \\
PromptStyle~\citep{liu2023promptstyle}                  & --   & --   & --   & 57.4 & 46.4 & 30.9 \\
Parler-TTS-large~\citep{lyth2024natural}               & --   & --   & --   & 60.0 & 45.9 & 31.2 \\
VoxInstruct~\citep{zhou2024voxinstruct}                  & 47.5 & 52.3 & 42.6 & 54.9 & 57.0 & 39.3 \\
VoiceSculptor~\citep{hu2026voicesculptor}                & 75.7 & 64.7 & 61.5 & --   & --   & --   \\
Mimo-Audio-7B-Instruct~\citep{zhang2025mimo}       & 75.7 & 74.3 & 61.5 & 80.6 & 77.6 & 59.5 \\
Qwen3-TTS-12Hz-1.7B-VD~\citep{hu2026qwen3}       & \textbf{85.2} & \textbf{81.1} & 65.1 & 82.9 & 82.4 & 68.4 \\
MOSS-VoiceGenerator~\citep{huang2026moss}             & 78.0 & 80.0 & \textbf{74.0} & 68.2 & 82.0 & 68.7 \\
\textbf{\modelname{}}        & \textbf{85.2} & 71.5 & 60.8 & \textbf{84.2} & \textbf{83.2} & \textbf{71.4} \\
\bottomrule
\end{tabular}
\end{table}

\subsection{AudioVAE V2 Reconstruction Quality}
\label{sec:exp-audiovae}

\begin{table}[t]
\centering
\small
\caption{Reconstruction quality of the Audio VAE used in VoxCPM, VoxCPM1.5, and VoxCPM2. Evaluation metrics: MelD-48k: Mel-distance at 48kHz; MelD-16k: Mel-distance at 16kHz; PESQ-16k: Perceptual Speech Quality (Evaluated at 16kHz); STOI-16k: Short-Time Objective Intelligibility (Evaluated at 16kHz).}
\label{tab:audiovae}
\setlength{\tabcolsep}{4pt}
{\footnotesize
\begin{tabular}{lllcccccc}
\toprule
\multirow{2}{1cm}{\textbf{VAE Model}} & \multicolumn{2}{c}{\textbf{Sample Rate}} & \multicolumn{4}{c}{\textbf{VCTK (48kHz)}} & \multicolumn{2}{c}{\textbf{Song Describer (44kHz)}} \\
\cmidrule(lr){2-3}\cmidrule(lr){4-7} \cmidrule(lr){8-9}
& \textbf{Input} & \textbf{Output} & \textbf{MelD-48k}$\downarrow$ & \textbf{MelD-16k}$\downarrow$ & \textbf{STOI-16k}$\uparrow$ & \textbf{PESQ-16k}$\uparrow$ & \textbf{MelD-48k}$\downarrow$ & \textbf{MelD-16k}$\downarrow$  \\
\midrule
VoxCPM & 16kHz & 16kHz & 1.787 & \textbf{0.801} & \textbf{0.911} & \textbf{3.940} & 2.371 & \underline{1.246}  \\
VoxCPM1.5 & 44kHz & 44kHz & \textbf{1.139}  & 0.926 & 0.836 & 3.148 & \textbf{1.267} & 1.311  \\
\textbf{VoxCPM2} & \textbf{16kHz} & \textbf{48kHz} & \underline{1.335} & \underline{0.813} & \underline{0.907} & \underline{3.906} & \underline{1.334} & \textbf{1.133} \\
\bottomrule
\end{tabular}
}
\end{table}

As stated in Section~3.2, AudioVAE~V2 defines the latent interface used by the rest of the system.
The reconstruction fidelity of AudioVAE V2 therefore provides an approximate upper bound on the acoustic fidelity that the downstream generation pipeline can achieve under perfect latent prediction.
Here we report the reconstruction quality of the Audio VAE V2 used VoxCPM2.

Table~\ref{tab:audiovae} compares the Audio VAE used in VoxCPM, VoxCPM1.5, and AudioVAE V2 in \modelname{}
on held-out speech (VCTK) and singing (Song Describer) reconstruction sets.
The models differ in both input and output sample rate: VoxCPM reconstructs 16~kHz audio,
VoxCPM1.5 operates at 44~kHz, and \modelname{} uses a 16~kHz encoder with a 48~kHz decoder.
We report Mel-distance at both full-band and 16~kHz bandwidths,
together with 16~kHz speech quality (PESQ) and intelligibility metrics (STOI) on VCTK.

The comparison reveals complementary strengths across the three codec generations.
The Audio VAE used in VoxCPM remains highly competitive on 16~kHz-band speech metrics,
while the Audio VAE used in VoxCPM1.5 achieves the best full-band Mel-distance by operating directly at a high sample rate.
In contrast, AudioVAE V2 delivers competitive reconstruction across both low- and full-band metrics under a more challenging super-resolution setting:
it encodes 16~kHz audio but reconstructs at 48~kHz.
This result supports the asymmetric codec design introduced in Section~3.2,
which gives \modelname{} greater flexibility by keeping the autoregressive backbone in a compact low-rate latent space
while still enabling high-sample-rate waveform generation.

\subsection{Inference Efficiency and Deployment}
\label{sec:exp-efficiency}

\begin{table}[t]
\centering
\small
\caption{Inference efficiency of \modelname{} on a single NVIDIA RTX~4090 (24~GB). RTF (real-time factor) is wall-clock generation time divided by output-audio duration; values $<$1 indicate faster-than-real-time generation.}
\label{tab:efficiency}
\setlength{\tabcolsep}{8pt}
\begin{tabular}{lccc}
\toprule
\textbf{Inference path} & \textbf{Params} & \textbf{RTF $\downarrow$} & \textbf{VRAM} \\
\midrule
\modelname{} (PyTorch)        & 2B & 0.30 & $\sim$8\,GB \\
\modelname{} (Nano-vLLM)        & 2B & \textbf{0.13} & $\sim$8\,GB \\
\midrule
VoxCPM1.5 (PyTorch)  & 0.8B & 0.15 & $\sim$6\,GB \\
VoxCPM (PyTorch) & 0.6B & 0.17 & $\sim$5\,GB \\
\bottomrule
\end{tabular}
\end{table}

Practical usability of a multilingual and controllable TTS foundation model depends heavily on inference latency and deployment cost. We therefore evaluate the runtime efficiency of \modelname{} on a single NVIDIA RTX 4090 GPU (24 GB). Table~\ref{tab:efficiency} summarizes the results under different serving paths.

\paragraph{Latency and memory.}
Under the standard PyTorch implementation, \modelname{} achieves an RTF of 0.30 while consuming approximately 8 GB of peak VRAM, which is well within the capability of consumer-grade GPUs. With the optimized Nano-vLLM\footnote{\url{https://github.com/a710128/nanovllm-voxcpm}} serving engine, the RTF improves to 0.13, delivering over 7× real-time generation speed on the same hardware. Despite having 2.5× more parameters than VoxCPM1.5, \modelname{} maintains competitive or better RTF under optimized serving paths. This efficiency stems largely from retaining the compact 6.25 Hz token rate on the language-model side.

\begin{table}[t]
\centering
\small
\caption{Subjective evaluation on zero-shot voice cloning. N-MOS and S-MOS are reported with 95\% confidence intervals.}
\label{tab:mos_zeroshot}
\setlength{\tabcolsep}{8pt}
\begin{tabular}{lcc}
\toprule
\textbf{System} & \textbf{N-MOS}$\uparrow$ & \textbf{S-MOS}$\uparrow$ \\
\midrule
IndexTTS2                        & $4.78\pm0.02$ & $4.71\pm0.03$ \\
Qwen3-TTS                          & $4.80\pm0.02$ & $4.69\pm0.03$ \\
Fish Audio S2                       & $4.77\pm0.02$ & $4.69\pm0.03$ \\
LongCat-Audio-DiT                  & $4.63\pm0.03$ & $4.65\pm0.03$ \\
\textbf{\modelname{}}              & $4.78\pm0.02$ & $4.74\pm0.03$ \\
\bottomrule
\end{tabular}
\end{table}

\begin{table}[t]
\centering
\small
\caption{Subjective evaluation on multilingual synthesis. N-MOS averaged across eight languages which all systems support (de, en, es, fr, ja, ko, ru, zh) from MiniMax-MLS-Test.}
\label{tab:mos_multilingual}
\setlength{\tabcolsep}{8pt}
\begin{tabular}{lcc}
\toprule
\textbf{System} & \textbf{N-MOS}$\uparrow$ & \textbf{S-MOS}$\uparrow$ \\
\midrule
OmniVoice           & $4.76\pm0.02$ & $4.72\pm0.02$ \\
Qwen3-TTS            & $4.77\pm0.02$ & $4.60\pm0.03$ \\
Fish Audio S2          & $4.76\pm0.02$ & $4.70\pm0.03$ \\
\textbf{\modelname{}} & $4.78\pm0.02$ & $4.66\pm0.03$ \\
\bottomrule
\end{tabular}
\end{table}

\begin{table}[t]
\centering
\small
\caption{Subjective results on controllable generation. I-MOS rates how well the speech follows the text-based voice design instructions.}
\label{tab:mos_control}
\setlength{\tabcolsep}{8pt}
\begin{tabular}{lcc}
\toprule
\textbf{System} & \textbf{N-MOS}$\uparrow$ & \textbf{I-MOS}$\uparrow$ \\
\midrule
VoiceSculptor        & $3.69\pm0.07$ & $3.56\pm0.07$ \\
Qwen3-TTS-VD        & $4.61\pm0.03$ & $4.41\pm0.04$ \\
MOSS-VoiceGenerator          & $4.31\pm0.05$ & $4.15\pm0.05$ \\
\textbf{\modelname{}}         & $4.48\pm0.04$ & $4.50\pm0.03$ \\
\bottomrule
\end{tabular}
\end{table}

\paragraph{Streaming and Production Serving.}
The causal design of the TSLM and RALM, together with the patch-local nature of the LocDiT, naturally supports chunk-based streaming inference. This enables low-latency interactive applications. For high-throughput production environments, \modelname{} is also compatible with vLLM-Omni\footnote{\url{https://github.com/vllm-project/vllm-omni}}, allowing efficient batched serving. 
For detailed benchmarks regarding concurrency, throughput, and real-time factors, please refer to the repository documentation.

\subsection{Subjective Listening Tests}
\label{sec:exp-subjective}
To evaluate the perceived quality of TTS systems, we conduct subjective Mean Opinion Score (MOS) tests focusing on three key dimensions: naturalness (\textbf{N-MOS}), speaker similarity (\textbf{S-MOS}), and instruction-following (\textbf{I-MOS}), using a standard 5-point scale.

\paragraph{Setup.}
We prepare about 100 audio samples covering zero-shot voice cloning (English and Chinese from Seed-TTS-Eval), multilingual synthesis (8 languages from MiniMax-MLS-Test), and controllable generation with diverse natural-language instructions (written by LLM).
We compare \modelname{} against representative open-source baselines drawn from Sections~\ref{sec:exp-seedtts}--\ref{sec:exp-control}. 
A total of 50 listeners participated in a randomized, double-blind evaluation, and the three scoring metrics are defined as follows:
\begin{itemize}
    \item \textbf{N-MOS} (Naturalness MOS): Rates how natural and human-like the synthesized speech sounds.
    \item \textbf{S-MOS} (Similarity MOS): Rates how closely the voice matches the reference speaker, including timbre, accent and speaking styles.
    \item \textbf{I-MOS} (Instruction MOS): Rates how accurately the audio follows the text-based voice design instructions.
\end{itemize}

\paragraph{Results and Analysis.}
The baseline selections and MOS results are shown in Tables~\ref{tab:mos_zeroshot}, \ref{tab:mos_multilingual}, and~\ref{tab:mos_control}.
As shown in Table~\ref{tab:mos_zeroshot}, \modelname{} achieves strong performance in zero-shot cloning with N-MOS of 4.78 and the highest S-MOS of 4.74, demonstrating excellent naturalness and speaker fidelity. 
For multilingual synthesis (Table~\ref{tab:mos_multilingual}), \modelname{} attains the highest average N-MOS of 4.78 while maintaining competitive S-MOS (4.66) across eight languages.
In controllable generation (Table~\ref{tab:mos_control}), \modelname{} delivers competitive N-MOS (4.48) and the highest I-MOS (4.50), showing superior instruction adherence.
Overall, the subjective results are largely consistent with the trends observed in the objective metrics. Through our progressive training curriculum, \modelname{} successfully integrates multilingual, zero-shot cloning, and controllable capabilities within a single unified model, achieving performance comparable to or better than specialized systems across all tested dimensions. At the same time, these results also indicate that objective metrics such as speaker embedding cosine similarity (SIM) do not always fully reflect perceived cloning quality, as human listeners tend to be more sensitive to some fine-grained personal characteristics and speaking style consistency.

\section{Conclusion and Future Work}
\label{sec:conclusion}

In this work, we presented \modelname{}, a unified multilingual and controllable speech generation foundation model that extends the hierarchical continuous-latent paradigm of VoxCPM. 
Built on a single 2B-parameter backbone, \modelname{} natively supports 48~kHz synthesis across 30 languages and 9 Chinese dialects, along with multiple generation modes—including zero-shot voice cloning, natural-language voice design, and fine-grained style control—without relying on task-specific variants or external discrete tokenizers.
Extensive experiments on public and internal benchmarks demonstrate that our unified approach achieves competitive or state-of-the-art performance, offering a strong balance between scalability, voice fidelity, and controllability.

While \modelname{} delivers promising results, several challenges remain. 
Cross-lingual synthesis quality is still influenced by data distribution imbalances, resulting in performance variations on certain low-resource languages. 
Capturing highly abstract or complex instructions consistently across languages also presents difficulties, and further reducing computational and deployment overhead for large-scale serving remains an important direction.
Future work will focus on expanding high-quality long-tail corpora, refining the underlying continuous-latent representations, and improving instruction-following robustness across diverse speaking styles and languages. 
We have also introduced preliminary singing voice generation capability as an extension of controllable synthesis (see the project demo page for examples), although singing quality still has substantial room for improvement. 
In the longer term, we plan to explore unifying various audio generation tasks—speech, singing, and potentially other modalities—within the same hierarchical continuous-latent framework.

Finally, as high-fidelity speech generation technology becomes increasingly accessible, responsible deployment is essential. 
Future iterations will incorporate stronger safeguards such as content provenance tracking, digital watermarking, and voice cloning detection. 
We hope that the open release of \modelname{}, together with its training code and efficient adaptation tools under the Apache 2.0 license, provides a valuable and responsible foundation for both the research community and the broader open-source ecosystem.

\section{Contributors}
\modelname{} is a collaborative release by the Tsinghua Shenzhen International Graduate School (SIGS) Human-Computer Speech Interaction Lab (THUHCSI), Natural Language Processing Lab at Tsinghua University (THUNLP) and ModelBest. We would also like to thank the OpenBMB community for all their support.

\textbf{Core Contributors:} \quad
Yixuan Zhou, Guoyang Zeng, Xin Liu, Xiang Li, Renjie Yu, Jiancheng Gui, Jiaheng Wu, Ziyang Wang, Xudong Shen, Runchuan Ye, Zhisheng Zhang, Jiuyang Zhou, Bingsong Bai, Weiyue Sun, Mengyuan Deng, Qundong Shi, Zhiyong Wu, Zhiyuan Liu

\textbf{Other Contributors (Alphabetical order):} \quad
Biyuan Lin, Caixian Chen, Chao Jia, Chenzhe Jing, Daixi Zeng, Jiayi Zhang, Jie Zhou, Jilong Ma, Jie Sun, Ling Zheng, Minmin Fan, Siyuan Huang, Shuo Wang, Susu Bai, Wenxi Yang, YingJiao Wang, Yitong Wang, Zhen Luo, Zhizheng Yang, Zhong Zhuang

\newpage

\bibliographystyle{citation}
\bibliography{citation}

@inproceedings{ping2017deep,
  title={Deep Voice 3: Scaling Text-to-Speech with Convolutional Sequence Learning},
  author={Ping, Wei and Peng, Kainan and Gibiansky, Andrew and Arik, Sercan O and Kannan, Ajay and Narang, Sharan and Raiman, Jonathan and Miller, John},
  booktitle={International Conference on Learning Representations},
  year={2018}
}

@inproceedings{renfastspeech,
  title={FastSpeech 2: Fast and High-Quality End-to-End Text to Speech},
  author={Ren, Yi and Hu, Chenxu and Tan, Xu and Qin, Tao and Zhao, Sheng and Zhao, Zhou and Liu, Tie-Yan},
  booktitle={International Conference on Learning Representations},
  year={2020}
}

@inproceedings{li2019neural,
  title={Neural speech synthesis with transformer network},
  author={Li, Naihan and Liu, Shujie and Liu, Yanqing and Zhao, Sheng and Liu, Ming},
  booktitle={Proceedings of the AAAI conference on artificial intelligence},
  volume={33},
  number={01},
  pages={6706--6713},
  year={2019}
}

@inproceedings{wang2018style,
  title={Style tokens: Unsupervised style modeling, control and transfer in end-to-end speech synthesis},
  author={Wang, Yuxuan and Stanton, Daisy and Zhang, Yu and Ryan, RJ-Skerry and Battenberg, Eric and Shor, Joel and Xiao, Ying and Jia, Ye and Ren, Fei and Saurous, Rif A},
  booktitle={International conference on machine learning},
  pages={5180--5189},
  year={2018},
  organization={PMLR}
}

@inproceedings{cai2021emotion,
  title={Emotion controllable speech synthesis using emotion-unlabeled dataset with the assistance of cross-domain speech emotion recognition},
  author={Cai, Xiong and Dai, Dongyang and Wu, Zhiyong and Li, Xiang and Li, Jingbei and Meng, Helen},
  booktitle={ICASSP 2021-2021 IEEE International Conference on Acoustics, Speech and Signal Processing (ICASSP)},
  pages={5734--5738},
  year={2021},
  organization={IEEE}
}

@article{chen2025neural,
  title={Neural codec language models are zero-shot text to speech synthesizers},
  author={Chen, Sanyuan and Wang, Chengyi and Wu, Yu and Zhang, Ziqiang and Zhou, Long and Liu, Shujie and Chen, Zhuo and Liu, Yanqing and Wang, Huaming and Li, Jinyu and others},
  journal={IEEE Transactions on Audio, Speech and Language Processing},
  year={2025},
  publisher={IEEE}
}

@article{borsos2023audiolm,
  title={Audiolm: a language modeling approach to audio generation},
  author={Borsos, Zal{\'a}n and Marinier, Rapha{\"e}l and Vincent, Damien and Kharitonov, Eugene and Pietquin, Olivier and Sharifi, Matt and Roblek, Dominik and Teboul, Olivier and Grangier, David and Tagliasacchi, Marco and others},
  journal={IEEE/ACM transactions on audio, speech, and language processing},
  volume={31},
  pages={2523--2533},
  year={2023},
  publisher={IEEE}
}

@article{kharitonov2023speak,
  title={Speak, read and prompt: High-fidelity text-to-speech with minimal supervision},
  author={Kharitonov, Eugene and Vincent, Damien and Borsos, Zal{\'a}n and Marinier, Rapha{\"e}l and Girgin, Sertan and Pietquin, Olivier and Sharifi, Matt and Tagliasacchi, Marco and Zeghidour, Neil},
  journal={Transactions of the Association for Computational Linguistics},
  volume={11},
  pages={1703--1718},
  year={2023},
  publisher={MIT Press One Broadway, 12th Floor, Cambridge, Massachusetts 02142, USA~…}
}

@article{borsos2023soundstorm,
  title={Soundstorm: Efficient parallel audio generation},
  author={Borsos, Zal{\'a}n and Sharifi, Matt and Vincent, Damien and Kharitonov, Eugene and Zeghidour, Neil and Tagliasacchi, Marco},
  journal={arXiv preprint arXiv:2305.09636},
  year={2023}
}

@article{defossez2022high,
  title={High fidelity neural audio compression},
  author={D{\'e}fossez, Alexandre and Copet, Jade and Synnaeve, Gabriel and Adi, Yossi},
  journal={arXiv preprint arXiv:2210.13438},
  year={2022}
}

@article{kumar2023high,
  title={High-fidelity audio compression with improved rvqgan},
  author={Kumar, Rithesh and Seetharaman, Prem and Luebs, Alejandro and Kumar, Ishaan and Kumar, Kundan},
  journal={Advances in Neural Information Processing Systems},
  volume={36},
  pages={27980--27993},
  year={2023}
}

@article{xin2024bigcodec,
  title={Bigcodec: Pushing the limits of low-bitrate neural speech codec},
  author={Xin, Detai and Tan, Xu and Takamichi, Shinnosuke and Saruwatari, Hiroshi},
  journal={arXiv preprint arXiv:2409.05377},
  year={2024}
}

@article{casanova2024xtts,
  title={Xtts: a massively multilingual zero-shot text-to-speech model},
  author={Casanova, Edresson and Davis, Kelly and G{\"o}lge, Eren and G{\"o}knar, G{\"o}rkem and Gulea, Iulian and Hart, Logan and Aljafari, Aya and Meyer, Joshua and Morais, Reuben and Olayemi, Samuel and others},
  journal={arXiv preprint arXiv:2406.04904},
  year={2024}
}

@inproceedings{wangmaskgct,
  title={MaskGCT: Zero-Shot Text-to-Speech with Masked Generative Codec Transformer},
  author={Wang, Yuancheng and Zhan, Haoyue and Liu, Liwei and Zeng, Ruihong and Guo, Haotian and Zheng, Jiachen and Zhang, Qiang and Zhang, Xueyao and Zhang, Shunsi and Wu, Zhizheng},
  year={2025},
  booktitle={The Thirteenth International Conference on Learning Representations}
}

@inproceedings{peng2024voicecraft,
  title={VoiceCraft: Zero-Shot Speech Editing and Text-to-Speech in the Wild},
  author={Peng, Puyuan and Huang, Po-Yao and Li, Shang-Wen and Mohamed, Abdelrahman and Harwath, David},
  booktitle={Proceedings of the 62nd Annual Meeting of the Association for Computational Linguistics (Volume 1: Long Papers)},
  pages={12442--12462},
  year={2024}
}

@inproceedings{shen2023naturalspeech,
  title={NaturalSpeech 2: Latent Diffusion Models are Natural and Zero-Shot Speech and Singing Synthesizers},
  author={Shen, Kai and Ju, Zeqian and Tan, Xu and Liu, Eric and Leng, Yichong and He, Lei and Qin, Tao and Bian, Jiang and others},
  booktitle={The Twelfth International Conference on Learning Representations},
  year={2023}
}

@inproceedings{zhang2024speechtokenizer,
  title={Speechtokenizer: Unified speech tokenizer for speech language models},
  author={Zhang, Xin and Zhang, Dong and Li, Shimin and Zhou, Yaqian and Qiu, Xipeng},
  booktitle={International Conference on Learning Representations},
  volume={2024},
  pages={31798--31818},
  year={2024}
}

@inproceedings{chen2024f5,
  title={F5-tts: A fairytaler that fakes fluent and faithful speech with flow matching},
  author={Chen, Yushen and Niu, Zhikang and Ma, Ziyang and Deng, Keqi and Wang, Chunhui and JianZhao, JianZhao and Yu, Kai and Chen, Xie},
  booktitle={Proceedings of the 63rd Annual Meeting of the Association for Computational Linguistics (Volume 1: Long Papers)},
  pages={6255--6271},
  year={2025}
}

@article{le2023voicebox,
  title={Voicebox: Text-guided multilingual universal speech generation at scale},
  author={Le, Matthew and Vyas, Apoorv and Shi, Bowen and Karrer, Brian and Sari, Leda and Moritz, Rashel and Williamson, Mary and Manohar, Vimal and Adi, Yossi and Mahadeokar, Jay and others},
  journal={Advances in neural information processing systems},
  volume={36},
  pages={14005--14034},
  year={2023}
}

@article{du2024cosyvoice1,
  title={Cosyvoice: A scalable multilingual zero-shot text-to-speech synthesizer based on supervised semantic tokens},
  author={Du, Zhihao and Chen, Qian and Zhang, Shiliang and Hu, Kai and Lu, Heng and Yang, Yexin and Hu, Hangrui and Zheng, Siqi and Gu, Yue and Ma, Ziyang and others},
  journal={arXiv preprint arXiv:2407.05407},
  year={2024}
}

@article{du2024cosyvoice2,
  title={Cosyvoice 2: Scalable streaming speech synthesis with large language models},
  author={Du, Zhihao and Wang, Yuxuan and Chen, Qian and Shi, Xian and Lv, Xiang and Zhao, Tianyu and Gao, Zhifu and Yang, Yexin and Gao, Changfeng and Wang, Hui and others},
  journal={arXiv preprint arXiv:2412.10117},
  year={2024}
}

@article{du2025cosyvoice3,
  title={Cosyvoice 3: Towards in-the-wild speech generation via scaling-up and post-training},
  author={Du, Zhihao and Gao, Changfeng and Wang, Yuxuan and Yu, Fan and Zhao, Tianyu and Wang, Hao and Lv, Xiang and Wang, Hui and Ni, Chongjia and Shi, Xian and others},
  journal={arXiv preprint arXiv:2505.17589},
  year={2025}
}

@inproceedings{zhou2025indextts2,
  title={Indextts2: A breakthrough in emotionally expressive and duration-controlled auto-regressive zero-shot text-to-speech},
  author={Zhou, Siyi and Zhou, Yiquan and He, Yi and Zhou, Xun and Wang, Jinchao and Deng, Wei and Shu, Jingchen},
  booktitle={Proceedings of the AAAI Conference on Artificial Intelligence},
  volume={40},
  number={41},
  pages={35139--35148},
  year={2026}
}

@article{liao2026fish,
  title={Fish Audio S2 Technical Report},
  author={Liao, Shijia and Wang, Yuxuan and Liu, Songting and Cheng, Yifan and Zhang, Ruoyi and Li, Tianyu and Li, Shidong and Zheng, Yisheng and Liu, Xingwei and Wang, Qingzheng and others},
  journal={arXiv preprint arXiv:2603.08823},
  year={2026}
}

@article{ye2025llasa,
  title={Llasa: Scaling train-time and inference-time compute for llama-based speech synthesis},
  author={Ye, Zhen and Zhu, Xinfa and Chan, Chi-Min and Wang, Xinsheng and Tan, Xu and Lei, Jiahe and Peng, Yi and Liu, Haohe and Jin, Yizhu and Dai, Zheqi and others},
  journal={arXiv preprint arXiv:2502.04128},
  year={2025}
}

@article{gong2026moss,
  title={MOSS-TTS Technical Report},
  author={Gong, Yitian and Jiang, Botian and Zhao, Yiwei and Yuan, Yucheng and Chen, Kuangwei and Jiang, Yaozhou and Chang, Cheng and Hong, Dong and Chen, Mingshu and Li, Ruixiao and others},
  journal={arXiv preprint arXiv:2603.18090},
  year={2026}
}

@article{zhu2026omnivoice,
  title={OmniVoice: Towards Omnilingual Zero-Shot Text-to-Speech with Diffusion Language Models},
  author={Zhu, Han and Ye, Lingxuan and Kang, Wei and Yao, Zengwei and Guo, Liyong and Kuang, Fangjun and Han, Zhifeng and Zhuang, Weiji and Lin, Long and Povey, Daniel},
  journal={arXiv preprint arXiv:2604.00688},
  year={2026}
}

@article{hu2026qwen3,
  title={Qwen3-TTS Technical Report},
  author={Hu, Hangrui and Zhu, Xinfa and He, Ting and Guo, Dake and Zhang, Bin and Wang, Xiong and Guo, Zhifang and Jiang, Ziyue and Hao, Hongkun and Guo, Zishan and others},
  journal={arXiv preprint arXiv:2601.15621},
  year={2026}
}

@article{liu2026voxtral,
  title={Voxtral TTS},
  author={Liu, Alexander H and Tacnet, Alexis and Ehrenberg, Andy and Lo, Andy and Sun, Chen-Yo and Lample, Guillaume and Lagarde, Henry and Delignon, Jean-Malo and Kim, Jaeyoung and Harvill, John and others},
  journal={arXiv preprint arXiv:2603.25551},
  year={2026}
}

@inproceedings{eskimez2024e2,
  title={E2 tts: Embarrassingly easy fully non-autoregressive zero-shot tts},
  author={Eskimez, Sefik Emre and Wang, Xiaofei and Thakker, Manthan and Li, Canrun and Tsai, Chung-Hsien and Xiao, Zhen and Yang, Hemin and Zhu, Zirun and Tang, Min and Tan, Xu and others},
  booktitle={2024 IEEE spoken language technology workshop (SLT)},
  pages={682--689},
  year={2024},
  organization={IEEE}
}

@inproceedings{shen2018natural,
  title={Natural tts synthesis by conditioning wavenet on mel spectrogram predictions},
  author={Shen, Jonathan and Pang, Ruoming and Weiss, Ron J and Schuster, Mike and Jaitly, Navdeep and Yang, Zongheng and Chen, Zhifeng and Zhang, Yu and Wang, Yuxuan and Skerrv-Ryan, Rj and others},
  booktitle={2018 IEEE international conference on acoustics, speech and signal processing (ICASSP)},
  pages={4779--4783},
  year={2018},
  organization={IEEE}
}

@inproceedings{meng2024autoregressive,
  title={Autoregressive speech synthesis without vector quantization},
  author={Meng, Lingwei and Zhou, Long and Liu, Shujie and Chen, Sanyuan and Han, Bing and Hu, Shujie and Liu, Yanqing and Li, Jinyu and Zhao, Sheng and Wu, Xixin and others},
  booktitle={Proceedings of the 63rd Annual Meeting of the Association for Computational Linguistics (Volume 1: Long Papers)},
  pages={1287--1300},
  year={2025}
}

@article{li2024autoregressive,
  title={Autoregressive image generation without vector quantization},
  author={Li, Tianhong and Tian, Yonglong and Li, He and Deng, Mingyang and He, Kaiming},
  journal={Advances in Neural Information Processing Systems},
  volume={37},
  pages={56424--56445},
  year={2024}
}

@inproceedings{jia2025ditar,
  title={DiTAR: Diffusion Transformer Autoregressive Modeling for Speech Generation},
  author={Jia, Dongya and Chen, Zhuo and Chen, Jiawei and Du, Chenpeng and Wu, Jian and Cong, Jian and Zhuang, Xiaobin and Li, Chumin and Wei, Zhen and Wang, Yuping and others},
  booktitle={International Conference on Machine Learning},
  pages={27255--27270},
  year={2025},
  organization={PMLR}
}

@article{peng2025vibevoice,
  title={Vibevoice technical report},
  author={Peng, Zhiliang and Yu, Jianwei and Wang, Wenhui and Chang, Yaoyao and Sun, Yutao and Dong, Li and Zhu, Yi and Xu, Weijiang and Bao, Hangbo and Wang, Zehua and others},
  journal={arXiv preprint arXiv:2508.19205},
  year={2025}
}

@inproceedings{turetzky2025speech,
  title={Speech Synthesis From Continuous Features Using Per-Token Latent Diffusion},
  author={Turetzky, Arnon and Dekel, Avihu and Shabtay, Nimrod and Shechtman, Slava and Haws, David and Aronowitz, Hagai and Hoory, Ron and Adi, Yossi},
  booktitle={2025 IEEE Automatic Speech Recognition and Understanding Workshop (ASRU)},
  pages={1--8},
  year={2025},
  organization={IEEE}
}

@inproceedings{an2026mela,
  title={Mela-tts: Joint transformer-diffusion model with representation alignment for speech synthesis},
  author={An, Keyu and Zhang, Zhiyu and Gao, Changfeng and Li, Yabin and Peng, Zhendong and Wang, Haoxu and Du, Zhihao and Zhao, Han and Gao, Zhifu and Li, Xiangang},
  booktitle={ICASSP 2026-2026 IEEE International Conference on Acoustics, Speech and Signal Processing (ICASSP)},
  pages={18337--18341},
  year={2026},
  organization={IEEE}
}

@inproceedings{wang2025felle,
  title={Felle: Autoregressive speech synthesis with token-wise coarse-to-fine flow matching},
  author={Wang, Hui and Liu, Shujie and Meng, Lingwei and Li, Jinyu and Yang, Yifan and Zhao, Shiwan and Sun, Haiyang and Liu, Yanqing and Sun, Haoqin and Zhou, Jiaming and others},
  booktitle={Proceedings of the 33rd ACM International Conference on Multimedia},
  pages={10229--10238},
  year={2025}
}

@article{wu2025clear,
  title={CLEAR: Continuous Latent Autoregressive Modeling for High-quality and Low-latency Speech Synthesis},
  author={Wu, Chun Yat and Deng, Jiajun and Li, Guinan and Kong, Qiuqiang and Lui, Simon},
  journal={arXiv preprint arXiv:2508.19098},
  year={2025}
}

@article{team2025minicpm4,
  title={Minicpm4: Ultra-efficient llms on end devices},
  author={Team, MiniCPM and Xiao, Chaojun and Li, Yuxuan and Han, Xu and Bai, Yuzhuo and Cai, Jie and Chen, Haotian and Chen, Wentong and Cong, Xin and Cui, Ganqu and others},
  journal={arXiv preprint arXiv:2506.07900},
  year={2025}
}

@inproceedings{mentzerfinite,
  title={Finite Scalar Quantization: VQ-VAE Made Simple},
  author={Mentzer, Fabian and Minnen, David and Agustsson, Eirikur and Tschannen, Michael},
  year={2024},
  booktitle={The Twelfth International Conference on Learning Representations}
}

@article{anastassiou2024seed,
  title={Seed-tts: A family of high-quality versatile speech generation models},
  author={Anastassiou, Philip and Chen, Jiawei and Chen, Jitong and Chen, Yuanzhe and Chen, Zhuo and Chen, Ziyi and Cong, Jian and Deng, Lelai and Ding, Chuang and Gao, Lu and others},
  journal={arXiv preprint arXiv:2406.02430},
  year={2024}
}

@article{zhang2025minimax,
  title={Minimax-speech: Intrinsic zero-shot text-to-speech with a learnable speaker encoder},
  author={Zhang, Bowen and Guo, Congchao and Yang, Geng and Yu, Hang and Zhang, Haozhe and Lei, Heidi and Mai, Jialong and Yan, Junjie and Yang, Kaiyue and Yang, Mingqi and others},
  journal={arXiv preprint arXiv:2505.07916},
  year={2025}
}

@article{guo2024fireredtts,
  title={Fireredtts: A foundation text-to-speech framework for industry-level generative speech applications},
  author={Guo, Hao-Han and Hu, Yao and Liu, Kun and Shen, Fei-Yu and Tang, Xu and Wu, Yi-Chen and Xie, Feng-Long and Xie, Kun and Xu, Kai-Tuo},
  journal={arXiv preprint arXiv:2409.03283},
  year={2024}
}

@article{xie2025fireredtts,
  title={FireRedTTS-2: Towards Long Conversational Speech Generation for Podcast and Chatbot},
  author={Xie, Kun and Shen, Feiyu and Li, Junjie and Xie, Fenglong and Tang, Xu and Hu, Yao},
  journal={arXiv preprint arXiv:2509.02020},
  year={2025}
}

@article{wang2025spark,
  title={Spark-tts: An efficient llm-based text-to-speech model with single-stream decoupled speech tokens},
  author={Wang, Xinsheng and Jiang, Mingqi and Ma, Ziyang and Zhang, Ziyu and Liu, Songxiang and Li, Linqin and Liang, Zheng and Zheng, Qixi and Wang, Rui and Feng, Xiaoqin and others},
  journal={arXiv preprint arXiv:2503.01710},
  year={2025}
}

@article{xu2025qwen2,
  title={Qwen2. 5-omni technical report},
  author={Xu, Jin and Guo, Zhifang and He, Jinzheng and Hu, Hangrui and He, Ting and Bai, Shuai and Chen, Keqin and Wang, Jialin and Fan, Yang and Dang, Kai and others},
  journal={arXiv preprint arXiv:2503.20215},
  year={2025}
}

@article{xu2025qwen3,
  title={Qwen3-omni technical report},
  author={Xu, Jin and Guo, Zhifang and Hu, Hangrui and Chu, Yunfei and Wang, Xiong and He, Jinzheng and Wang, Yuxuan and Shi, Xian and He, Ting and Zhu, Xinfa and others},
  journal={arXiv preprint arXiv:2509.17765},
  year={2025}
}

@inproceedings{baas2025mars6,
  title={MARS6: A Small and Robust Hierarchical-Codec Text-to-Speech Model},
  author={Baas, Matthew and Scholtz, Pieter and Mehta, Arnav and Dyson, Elliott and Prakash, Akshat and Kamper, Herman},
  booktitle={ICASSP 2025-2025 IEEE International Conference on Acoustics, Speech and Signal Processing (ICASSP)},
  pages={1--5},
  year={2025},
  organization={IEEE}
}

@article{xin2026longcat,
  title={LongCat-AudioDiT: High-Fidelity Diffusion Text-to-Speech in the Waveform Latent Space},
  author={Xin, Detai and Hu, Shujie and Yang, Chengzuo and Huang, Chen and Yu, Guoqiao and Wan, Guanglu and Cai, Xunliang},
  journal={arXiv preprint arXiv:2603.29339},
  year={2026}
}

@inproceedings{guo2023prompttts,
  title={Prompttts: Controllable text-to-speech with text descriptions},
  author={Guo, Zhifang and Leng, Yichong and Wu, Yihan and Zhao, Sheng and Tan, Xu},
  booktitle={ICASSP 2023-2023 IEEE International Conference on Acoustics, Speech and Signal Processing (ICASSP)},
  pages={1--5},
  year={2023},
  organization={IEEE}
}

@inproceedings{leng2024prompttts,
  title={Prompttts 2: Describing and generating voices with text prompt},
  author={Leng, Yichong and Guo, Zhifang and Shen, Kai and Ju, Zeqian and Tan, Xu and Liu, Eric and Liu, Yufei and Yang, Dongchao and Song, Kaitao and He, Lei and others},
  booktitle={International Conference on Learning Representations},
  volume={2024},
  pages={57672--57688},
  year={2024}
}

@article{yang2024instructtts,
  title={Instructtts: Modelling expressive tts in discrete latent space with natural language style prompt},
  author={Yang, Dongchao and Liu, Songxiang and Huang, Rongjie and Weng, Chao and Meng, Helen},
  journal={IEEE/ACM Transactions on Audio, Speech, and Language Processing},
  volume={32},
  pages={2913--2925},
  year={2024},
  publisher={IEEE}
}

@inproceedings{liu2023promptstyle,
  title={PromptStyle: Controllable Style Transfer for Text-to-Speech with Natural Language Descriptions},
  author={Liu, Guanghou and Zhang, Yongmao and Lei, Yi and Chen, Yunlin and Wang, Rui and Xie, Lei and Li, Zhifei},
  booktitle={Proc. Interspeech 2023},
  pages={4888--4892},
  year={2023}
}

@article{kazemnejad2023impact,
  title={The impact of positional encoding on length generalization in transformers},
  author={Kazemnejad, Amirhossein and Padhi, Inkit and Natesan Ramamurthy, Karthikeyan and Das, Payel and Reddy, Siva},
  journal={Advances in Neural Information Processing Systems},
  volume={36},
  pages={24892--24928},
  year={2023}
}

@inproceedings{xie2025towards,
  title={Towards controllable speech synthesis in the era of large language models: A systematic survey},
  author={Xie, Tianxin and Rong, Yan and Zhang, Pengfei and Wang, Wenwu and Liu, Li},
  booktitle={Proceedings of the 2025 Conference on Empirical Methods in Natural Language Processing},
  pages={764--791},
  year={2025}
}

@article{manku2026emergenttts,
  title={Emergenttts-eval: Evaluating tts models on complex prosodic, expressiveness, and linguistic challenges using model-as-a-judge},
  author={Manku, Ruskin Raj and Tang, Yuzhi and Shi, Xingjian and Li, Mu and Smola, Alexander},
  journal={Advances in Neural Information Processing Systems},
  volume={38},
  year={2026}
}

@article{mai2026magic,
  title={MAGIC-TTS: Fine-Grained Controllable Speech Synthesis with Explicit Local Duration and Pause Control},
  author={Mai, Jialong and Xing, Xiaofen and Xu, Xiangmin},
  journal={arXiv preprint arXiv:2604.21164},
  year={2026}
}

@inproceedings{ji2024textrolspeech,
  title={Textrolspeech: A text style control speech corpus with codec language text-to-speech models},
  author={Ji, Shengpeng and Zuo, Jialong and Fang, Minghui and Jiang, Ziyue and Chen, Feiyang and Duan, Xinyu and Huai, Baoxing and Zhao, Zhou},
  booktitle={ICASSP 2024-2024 IEEE International Conference on Acoustics, Speech and Signal Processing (ICASSP)},
  pages={10301--10305},
  year={2024},
  organization={IEEE}
}

@inproceedings{jin2024speechcraft,
  title={Speechcraft: A fine-grained expressive speech dataset with natural language description},
  author={Jin, Zeyu and Jia, Jia and Wang, Qixin and Li, Kehan and Zhou, Shuoyi and Zhou, Songtao and Qin, Xiaoyu and Wu, Zhiyong},
  booktitle={Proceedings of the 32nd ACM International Conference on Multimedia},
  pages={1255--1264},
  year={2024}
}

@article{wang2025capspeech,
  title={Capspeech: Enabling downstream applications in style-captioned text-to-speech},
  author={Wang, Helin and Hai, Jiarui and Chong, Dading and Thakkar, Karan and Feng, Tiantian and Yang, Dongchao and Lee, Junhyeok and Thebaud, Thomas and Velazquez, Laureano Moro and Villalba, Jesus and others},
  journal={arXiv preprint arXiv:2506.02863},
  year={2025}
}

@article{lyth2024natural,
  title={Natural language guidance of high-fidelity text-to-speech with synthetic annotations},
  author={Lyth, Dan and King, Simon},
  journal={arXiv preprint arXiv:2402.01912},
  year={2024}
}

@article{ren2026ov,
  title={OV-InstructTTS: Towards Open-Vocabulary Instruct Text-to-Speech},
  author={Ren, Yong and Yi, Jiangyan and Tao, Jianhua and Sun, Haiyang and Wen, Zhengqi and Gu, Hao and Xu, Le and Bai, Ye},
  journal={arXiv preprint arXiv:2601.01459},
  year={2026}
}

@article{huang2026moss,
  title={MOSS-VoiceGenerator: Create Realistic Voices with Natural Language Descriptions},
  author={Huang, Kexin and Fan, Liwei and Jiang, Botian and Jiang, Yaozhou and Tu, Qian and Zhu, Jie and Zhang, Yuqian and Zhao, Yiwei and Yang, Chenchen and Fei, Zhaoye and others},
  journal={arXiv preprint arXiv:2603.28086},
  year={2026}
}

@inproceedings{yang2025emovoice,
  title={Emovoice: Llm-based emotional text-to-speech model with freestyle text prompting},
  author={Yang, Guanrou and Yang, Chen and Chen, Qian and Ma, Ziyang and Chen, Wenxi and Wang, Wen and Wang, Tianrui and Yang, Yifan and Niu, Zhikang and Liu, Wenrui and others},
  booktitle={Proceedings of the 33rd ACM International Conference on Multimedia},
  pages={10748--10757},
  year={2025}
}

@article{huang2025instructttseval,
  title={Instructttseval: Benchmarking complex natural-language instruction following in text-to-speech systems},
  author={Huang, Kexin and Tu, Qian and Fan, Liwei and Yang, Chenchen and Zhang, Dong and Li, Shimin and Fei, Zhaoye and Cheng, Qinyuan and Qiu, Xipeng},
  journal={arXiv preprint arXiv:2506.16381},
  year={2025}
}

@article{chen2026mint,
  title={MINT-Bench: A Comprehensive Multilingual Benchmark for Instruction-Following Text-to-Speech},
  author={Chen, Huakang and Hu, Jingbin and Xue, Liumeng and Zhan, Qirui and Li, Wenhao and Ma, Guobin and Xie, Hanke and Guo, Dake and Ma, Linhan and Jiang, Yuepeng and others},
  journal={arXiv preprint arXiv:2604.17958},
  year={2026}
}

@article{hu2026voicesculptor,
  title={Voicesculptor: Your voice, designed by you},
  author={Hu, Jingbin and Chen, Huakang and Ma, Linhan and Guo, Dake and Zhan, Qirui and Li, Wenhao and Zhang, Haoyu and Xia, Kangxiang and Zhang, Ziyu and Tian, Wenjie and others},
  journal={arXiv preprint arXiv:2601.10629},
  year={2026}
}

@article{chen2026flexivoice,
  title={FlexiVoice: Enabling Flexible Style Control in Zero-Shot TTS with Natural Language Instructions},
  author={Chen, Dekun and Zhang, Xueyao and Wang, Yuancheng and Dai, Kenan and Ma, Li and Wu, Zhizheng},
  journal={arXiv preprint arXiv:2601.04656},
  year={2026}
}

@inproceedings{zhou2024voxinstruct,
  title={Voxinstruct: Expressive human instruction-to-speech generation with unified multilingual codec language modelling},
  author={Zhou, Yixuan and Qin, Xiaoyu and Jin, Zeyu and Zhou, Shuoyi and Lei, Shun and Zhou, Songtao and Wu, Zhiyong and Jia, Jia},
  booktitle={Proceedings of the 32nd ACM International Conference on Multimedia},
  pages={554--563},
  year={2024}
}

@article{openaudios1,
  title={OpenAudio S1: a cutting-edge text-to-speech model that performs like voice actors},
  author={OpenAudio},
  journal={https://openaudio.com/blogs/s1},
  year={2024}
}

@article{zhang2025mimo,
  title={MiMo-Audio: Audio Language Models are Few-Shot Learners},
  author={Zhang, Dong and Wang, Gang and Xue, Jinlong and Fang, Kai and Zhao, Liang and Ma, Rui and Ren, Shuhuai and Liu, Shuo and Guo, Tao and Zhuang, Weiji and others},
  journal={arXiv preprint arXiv:2512.23808},
  year={2025}
}

@article{zhu2025zipvoice,
  title={Zipvoice: Fast and high-quality zero-shot text-to-speech with flow matching},
  author={Zhu, Han and Kang, Wei and Yao, Zengwei and Guo, Liyong and Kuang, Fangjun and Li, Zhaoqing and Zhuang, Weiji and Lin, Long and Povey, Daniel},
  journal={arXiv preprint arXiv:2506.13053},
  year={2025}
}

@article{zhou2025voxcpm,
  title={Voxcpm: Tokenizer-free TTS for context-aware speech generation and true-to-life voice cloning},
  author={Zhou, Yixuan and Zeng, Guoyang and Liu, Xin and Li, Xiang and Yu, Renjie and Wang, Ziyang and Ye, Runchuan and Sun, Weiyue and Gui, Jiancheng and Li, Kehan and others},
  journal={arXiv preprint arXiv:2509.24650},
  year={2025}
}

@inproceedings{zhou2026hierarchical,
  title={Hierarchical Semantic-Acoustic Modeling via Semi-Discrete Residual Representations for Expressive End-to-End Speech Synthesis},
  author={Zhou, Yixuan and Zeng, Guoyang and Liu, Xin and Li, Xiang and Yu, Renjie and Wang, Ziyang and Ye, Runchuan and Sun, Weiyue and Gui, Jiancheng and Li, Kehan and others},
  booktitle={The Fourteenth International Conference on Learning Representations},
  year={2026}
}

@article{fan2025cfgzerostar,
  title   = {CFG-Zero*: Improved Classifier-Free Guidance for Flow Matching Models},
  author  = {Fan, Weichen and Zheng, Amber Yijia and Zhu, Dejia and Ma, Yao and Liu, Nikola and Wang, Zhangyang and Liu, Dilin},
  journal = {arXiv preprint arXiv:2503.18886},
  year    = {2025},
}

@article{lee2025hierspeechpp,
  title   = {{HierSpeech++}: Bridging the Gap between Semantic and Acoustic Representation of Speech by Hierarchical Variational Inference for Zero-Shot Speech Synthesis},
  author  = {Lee, Sang-Hoon and Choi, Ha-Yeong and Kim, Seung-Bin and Lee, Seong-Whan},
  journal = {{IEEE} Transactions on Neural Networks and Learning Systems},
  year    = {2025}
}

@inproceedings{nishimura2025halle,
  title     = {{HALL-E}: Hierarchical Neural Codec Language Model for Minute-Long Zero-Shot Text-to-Speech Synthesis},
  author    = {Nishimura, Yuto and Hirose, Takumi and Ohi, Masanari and Nakayama, Hideki and Inoue, Nakamasa},
  booktitle = {International Conference on Learning Representations ({ICLR})},
  year      = {2025}
}

@misc{bosonai2025higgsaudio,
  title        = {{Higgs Audio v2}: Redefining Expressiveness in Audio Generation},
  author       = {{Boson AI}},
  howpublished = {\url{https://github.com/boson-ai/higgs-audio}},
  year         = {2025}
}

@article{huang2025stepaudio,
  title   = {{Step-Audio}: Unified Understanding and Generation in Intelligent Speech Interaction},
  author  = {Huang, Ailin and Wu, Boyong and Wang, Bruce and Yan, Chao and Hu, Chen and Feng, Chengli and others},
  journal = {arXiv preprint arXiv:2502.11946},
  year    = {2025}
}

@article{kimiaudio2025,
  title   = {{Kimi-Audio} Technical Report},
  author  = {{Kimi Team}},
  journal = {arXiv preprint arXiv:2504.18425},
  year    = {2025}
}

@article{zeng2024glm4voice,
  title   = {{GLM-4-Voice}: Towards Intelligent and Human-Like End-to-End Spoken Chatbot},
  author  = {Zeng, Aohan and Du, Zhengxiao and Liu, Mingdao and Wang, Kedong and Jiang, Shengmin and Zhao, Lei and Dong, Yuxiao and Tang, Jie},
  journal = {arXiv preprint arXiv:2412.02612},
  year    = {2024}
}

@inproceedings{lee2024voiceldm,
  title     = {{VoiceLDM}: Text-to-Speech with Environmental Context},
  author    = {Lee, Yeonghyeon and Yeon, Inmo and Nam, Juhan and Chung, Joon Son},
  booktitle = {{IEEE} International Conference on Acoustics, Speech and Signal Processing ({ICASSP})},
  year      = {2024}
}

@article{vyas2024audiobox,
  title   = {{Audiobox}: Unified Audio Generation with Natural Language Prompts},
  author  = {Vyas, Apoorv and Shi, Bowen and Le, Matthew and Tjandra, Andros and Wu, Yi-Chiao and Guo, Baishan and Zhang, Jiemin and Zhang, Xinyue and Adkins, Robert and Ngan, William and others},
  journal = {arXiv preprint arXiv:2312.15821},
  year    = {2024}
}

@inproceedings{minixhofer2024ttsds,
  title     = {{TTSDS}---Text-to-Speech Distribution Score},
  author    = {Minixhofer, Christoph and Klejch, Ond{\v{r}}ej and Bell, Peter},
  booktitle = {{IEEE} Spoken Language Technology Workshop ({SLT})},
  year      = {2024}
}

@inproceedings{copet2024musicgen,
  title     = {Simple and Controllable Music Generation},
  author    = {Copet, Jade and Kreuk, Felix and Gat, Itai and Remez, Tal and Kant, David and
               Synnaeve, Gabriel and Adi, Yossi and D{\'e}fossez, Alexandre},
  booktitle = {Advances in Neural Information Processing Systems ({NeurIPS})},
  year      = {2024},
  note      = {arXiv:2306.05284}
}

@inproceedings{ye2024xcodec,
  title={Codec does matter: Exploring the semantic shortcoming of codec for audio language model},
  author={Ye, Zhen and Sun, Peiwen and Lei, Jiahe and Lin, Hongzhan and Tan, Xu and Dai, Zheqi and Kong, Qiuqiang and Chen, Jianyi and Pan, Jiahao and Liu, Qifeng and others},
  booktitle={Proceedings of the AAAI Conference on Artificial Intelligence},
  volume={39},
  number={24},
  pages={25697--25705},
  year={2025}
}

@article{su2024roformer,
  title={Roformer: Enhanced transformer with rotary position embedding},
  author={Su, Jianlin and Ahmed, Murtadha and Lu, Yu and Pan, Shengfeng and Bo, Wen and Liu, Yunfeng},
  journal={Neurocomputing},
  volume={568},
  pages={127063},
  year={2024},
  publisher={Elsevier}
}

@article{tian2025step,
  title={Step-Audio-R1 Technical Report},
  author={Tian, Fei and Zhang, Xiangyu Tony and Zhang, Yuxin and Zhang, Haoyang and Li, Yuxin and Liu, Daijiao and Deng, Yayue and Wu, Donghang and Chen, Jun and Zhao, Liang and others},
  journal={arXiv preprint arXiv:2511.15848},
  year={2025}
}

@article{meituan2026longcataudiodit,
  title   = {{LongCat-AudioDiT}: High-Fidelity Diffusion Text-to-Speech in the Waveform Latent Space},
  author  = {{Meituan LongCat Team}},
  journal = {arXiv preprint arXiv:2603.29339},
  year    = {2026}
}

@article{jiang2025megatts3,
  title   = {{MegaTTS} 3: Sparse Alignment Enhanced Latent Diffusion Transformer for Zero-Shot Speech Synthesis},
  author  = {Jiang, Ziyue and Ren, Yi and Li, Ruiqi and Ji, Shengpeng and Zhang, Boyang and
             Ye, Zhenhui and Zhang, Chen and Bai, Jionghao and Yang, Xiaoda and Zuo, Jialong and
             Zhang, Yu and Liu, Rui and Yin, Xiang and Zhao, Zhou},
  journal = {arXiv preprint arXiv:2502.18924},
  year    = {2025}
}

@article{xiao2025densing,
  title={Densing law of llms},
  author={Xiao, Chaojun and Cai, Jie and Zhao, Weilin and Lin, Biyuan and Zeng, Guoyang and Zhou, Jie and Zheng, Zhi and Han, Xu and Liu, Zhiyuan and Sun, Maosong},
  journal={Nature Machine Intelligence},
  pages={1--11},
  year={2025},
  publisher={Nature Publishing Group UK London}
}


\end{document}